\documentclass[useAMS,usenatbib,usegraphicx]{mn2e}

\usepackage{times}
\usepackage[T1]{fontenc} 
\usepackage{aecompl}

\title[Light cone anisotropy in 21 cm signal from the epoch of reionization]{Light cone anisotropy in the 21 cm signal from the epoch of reionization}
\author[K. Zawada et al.]{Karolina Zawada$^{1}$\thanks{E-mail:
karolina@astro.uni.torun.pl (KZ)}, Beno\^it Semelin$^{2,3}$\thanks{E-mail:
benoit.semelin@obspm.fr (BS)}, Patrick Vonlanthen$^{2}$, Sunghye Baek$^{4}$, Yves Revaz$^{5}$\\
$^{1}$Toru\'n Centre for Astronomy, Faculty of Physics, Astronomy and Informatics, Nicolaus Copernicus University, Grudziadzka 5, 87-100 Toru\'n, Poland\\
$^{2}$LERMA, Observatoire de Paris, 61 Av. de l'Observatoire, 75014 Paris, France\\
$^{3}$Universit\'{e} Pierre et Marie Curie, 4 Place Jules Janssen, 92195 Meudon Cedex, France\\
$^{4}$Scuola Normale Superiore, Piazza dei Cavalieri 7, 56126 Pisa, Italy\\
$^{5}$Laboratoire d'Astrophysique, Ecole Polytechnique F\'{e}d\'{e}rale de Lausanne (EPFL), Switzerland}
\begin{document}

\date{Accepted 2014 January 6.  Received 2014 January 6; in original form 2013 July 31}

\pagerange{\pageref{firstpage}--\pageref{lastpage}} \pubyear{2002}

\maketitle

\label{firstpage}

\begin{abstract}
Using a suite of detailed numerical simulations we estimate the level of anisotropy generated by the time evolution along the light cone of the $21$~cm signal from the epoch of reionization. Our simulations include the physics necessary to model the signal during both the late emission regime and the early absorption regime, namely X-ray and Lyman-band
3D radiative transfer in addition to the usual dynamics and ionizing UV transfer. The signal is analysed using correlation functions perpendicular and parallel to
the line of sight. We reproduce general findings from previous theoretical studies: the overall amplitude of the correlations and the fact that the light cone anisotropy is visible only
on large scales (100 comoving Mpc). However, the detailed behaviour is different. We find that, at three different epochs, the amplitude of the correlations along and perpendicular to the line of sight differ from each other, indicating anisotropy. We show that these three epochs are associated with three events of the global reionization history: the overlap
of ionized bubbles, the onset of mild heating by X-rays in regions around the sources, and the onset of 
efficient Lyman-$\alpha$ coupling in regions around the sources.
We find that a $20 \times 20 \: \mathrm{deg}^2$ survey area may be necessary to mitigate sample variance when we use the directional correlation functions. On a $100$~Mpc (comoving) scale, we show 
that the light cone anisotropy dominates over the anisotropy generated by peculiar velocity gradients computed in the linear regime. By modelling instrumental noise and limited resolution, we find
that the anisotropy should be easily detectable by the Square Kilometre Array, assuming perfect foreground removal, the limiting factor being a large enough survey size. In the case
of the Low-Frequency Array for radio astronomy, it is likely that only one anisotropy episode (ionized bubble overlap) will fall in the observing frequency range. This episode will be detectable only if sample variance is much reduced (i.e. a larger than $20 \times 20 \: \mathrm{deg}^2$ survey, which is not presently planned).

\end{abstract}

\begin{keywords}
numerical simulation, reionization, early Universe, large scale structure.
\end{keywords}

\section{Introduction}

The epoch of reionization (EoR) extends from redshift $z=20$--$30$ down to redshift $6$. During this era, the cold and neutral 
intergalactic medium (IGM) is progressively ionized by the light of the first stars and galaxies. To this day we have very little
observational information about the state of the IGM during this process. The Gunn-Peterson absorption in high-redshift quasars (QSO) \citep[]{Gunn65}
apparently shows that by $z \sim 6$ the Universe is more than $99.9 \%$ reionized \citep{Fan06}, although the current observational
sample may not be statistically significant \citep[see][]{Mesinger10}. The optical depth of Thomson scattering on free 
electrons measured from 
cosmic microwave background (CMB) observations
 favours an extended
epoch of reionization 
 \citep[the best fit from combined Wilkinson Microwave Anisotropy Probe (WMAP) and Planck data is $\tau= 0.089 \pm 0.014$,][]{Planck13}. Over the next few years, $21$~cm observations of the neutral IGM are the most promising type of observations likely to improve our understanding of reionization.

A number of projects are currently under way to detect the $21$~cm signal from reionization. The reionization project at 
the Giant Metre-wave Radio Telescope (GMRT) has
published an upper limit of $(248 \,\mathrm{mK})^2$ for the power spectrum at wavenumber $k=0.5\,h$~Mpc$^{-1}$ at $z=8.7$ \citep{Paciga13}. A similar upper
limit, $(300 \,\mathrm{mK})^2$  for comoving wavenumber $k=0.046$ Mpc$^{-1}$  at $z=9.5$, was found by the Murchison Widefield Array (MWA) \citep{Dillon13}. A stronger upper limit was found by the Precision Array for Probing the Epoch of Re-ionization (PAPER): $(52 \,\mathrm{mK})^2$ for $k=0.11 h$ Mpc$^{-1}$ at $z=7.7$ \citep{Parsons13}. The PAPER constraint
implies that a small amount of heating occurs at $z > 7.7$ in the neutral IGM (a few K), since otherwise the signal would have been
detected. This is not surprising: the opposite would imply that almost no X-rays were produced  prior to these
redshifts. The Low-Frequency Array for radio astronomy (LOFAR)  is in operation but has not yet published its results. The primary goal of all the Square Kilometre Array (SKA) pathfinders is to
achieve a statistical detection of the signal in the form of its three-dimensional (3D) power spectrum. 

In the process these pathfinders will have to deal with foreground contanimation, which is a thousand or more times brighter than the signal. The power spectrum is easier to detect than the flux density, as each single value is a statistic over a large number of Fourier modes. The level of noise affecting each single observed visibility is much reduced in the process, but a lot of information is lost. The full 3D imaging of the signal will mostly require SKA capabilities \citep[see][for expected capabilities]{Mellema13} although some low-resolution imaging should be possible with LOFAR \citep{Zaroubi12}.

Numerical simulations are an invaluable tool to predict and interpret the upcoming observations. Since power spectrum measurements are
expected to come first, the 3D spherically averaged power spectrum is the first observable quantity that simulations have focussed on. In
the emission regime (probably covering most of the EoR), predictions are in broad agreement with one another \citep[e.g.][]{Iliev08b,Lidz08}. In the
absorption regime, early during the EoR, predictions are also available \citep{Santos08,Baek10}, but are very sensitive to unknowns,
such as the relative amounts of X-ray, ionizing and Lyman-band photons, which depend on the nature of the sources and their 
formation history. Other statistical quantities able to detect non-Gaussianities, such as the pixel distribution function or the skewness of the brightness temperature distribution have been studied \citep{Ciardi03, Mellema06b, Harker09, Ichikawa10, Baek10}. How to exploit
the information in the full 3D tomography, however, is a subject that has barely been mined yet \citep{Vonlanthen11,Datta12b, Majumdar12}, but will
likely receive  a lot of attention in the upcoming years.

Somewhere between the low-noise low-information spherically averaged power spectrum and the high-noise high-information 3D imaging, lies
the angular power spectrum or, equivalently, the angular correlation function. Indeed, several factors induce an anisotropy in the power spectrum in such a way that properties are different along the line of sight (LOS) and perpendicular to the LOS. The first effect comes
from peculiar velocity gradients along the LOS that enhance or dim the $21$~cm brightness temperature, also called the Kaiser effect \citep[]{Kaiser87}. 
\citet{Barkana05} showed how this anisotropy, cosmological in nature, could be separated in the linear regime from astrophysical 
isotropic contributions to the brightness temperature fluctuations such as ionization patchiness and spin temperature fluctuations. Since then, their method has been tested for robustness in the non-linear regime and further refined \citep{Lidz07,Mao12,Shapiro13,Jensen13}. There is a second source of anisotropy that may interfere with the Kaiser effect in extracting cosmological information. Since we
will observe a light cone, points farther away will be seen earlier in the history of the EoR when bubbles were smaller, 
the IGM less heated by X-rays, and the Wouthuysen-Field coupling weaker. This again introduces an anisotropy where the LOS is a special
direction. 

\citet{Barkana06} (hereafter BL) first attempted to quantify this anisotropy using a simple theoretical model. They found that the anisotropy occurs during the later part
of the reionization history, when the neutral fraction $x_{\mathrm{n}}$ is smaller than $0.5$, on scales larger than
$50$ comoving Mpc. They quantified it by comparing the two-point correlation functions along and perpendicular to the line of sight. In their Fig. 4, they found that the peaks of the correlation functions computed along and perpendicular to the LOS occur at different times. They also found that the anisotropy is stronger in the case of an earlier 
reionization history (Pop III stars), and that the amplitude of the peaks is about $100\%$ higher in the case of Pop III stars than for the case of Pop II stars. Their model makes a number of assumptions whose robustness
is difficult to evaluate. One of the key ingredients is an estimate of the 
average radius of ionized bubbles as a function of redshift, based mainly  on 
the average overdensity inside the bubble. This ignores the wide scatter in 
ionized region sizes that occurs especially during the overlap phase. 
\citet{Chardin12} found that the linear size of the largest ionized region at 
$x_{\mathrm{n}} = 0.5$ is $\sim 20$ times larger than the average ionized region size.
Moreover, BL found that the anisotropy occurs in late reionization, during the overlap, when collective effects are most important and the 
ionization field is most non-Gaussian, making simple analytic estimations
of correlation functions more problematic.
Their predictions have not been cross-checked with full numerical simulations 
before, mainly because they require very large simulation boxes, and thus are 
costly. However, note the work by \citet{Datta12} that examines the light cone effect on the isotropic power spectrum in a $163$ cMpc box.  In this work we run full radiative transfer simulations in a $400$ h $^{-1}$~cMpc box \footnote{Hereafter, ``cMpc'' refers to comoving Mpc.}, including both stars and X-ray sources, and Lyman-$\alpha$ 
radiation for a fully self-consistent modelling of the early EoR. As we 
will see, even though we do find anisotropies on similar scales, our 
quantitative predictions differ from BL's.

\section[]{Description of the simulation suite}

The final product of our simulation suite is a data cube containing $\delta T_\mathrm{b}$, the differential brightness temperature of the $21$~cm signal. This quantity can be computed as \citep[e.g][]{Mellema13}: 
\begin{eqnarray}
\nonumber \delta T_{\mathrm{b}} & = & 28.1 \ x_{\mathrm{n}} \ (1 + \delta) \left( \frac{1+z}{10} \right)^{1/2} \frac{T_{\mathrm{s}}-T_{\gamma}}{T_{\mathrm{s}}} {1 \over (1 + {1 \over H} {{\mathrm d}v \over {\mathrm d}r})}\\
&& \times \left( \frac{\Omega_{\mathrm{b}}}{0.042} \frac{h}{0.73} \right) \left( \frac{0.24}{\Omega_{\mathrm{m}}} \right)^{1/2} \left( \frac{1 - Y_{\mathrm{p}}}{1-0.248} \right) \mathrm{mK},
\label{equdtb}
\end{eqnarray}
where $x_{\mathrm{n}}$ is the neutral hydrogen fraction, $\delta$ is the local baryon overdensity, $T_{\mathrm{s}}$ is the hydrogen spin 
temperature, $T_\gamma$ is the CMB temperature at redshift $z$, ${{\mathrm d}v \over {\mathrm d}r}$ is the peculiar velocity gradient along the line of 
sight and $H$, $\Omega_{\mathrm{b}}$, $\Omega_{\mathrm{m}}$, $h$, and $Y_{\mathrm{p}}$ are the usual cosmological parameters in a Friedmann--Lema\^{\i}tre--Robertson--Walker (FLRW) cosmological model in which inhomogeneity is modelled perturbatively or by assuming the Newtonian limit \citep[e.g.,][]{Komatsu11}. $T_{\mathrm s}$ itself, defined by the relative populations of the hyperfine level of the ground state of hydrogen, is the result of three 
competing processes: interactions with CMB photons drive $T_{\mathrm s}$ to $T_{\mathrm{CMB}}$, while collisions with other atoms or 
particles and interactions with Lyman-$\alpha$ photons \citep[the Wouthuysen-Field effect,][]{Wouthuysen52, Field58} drive it towards the local IGM gas temperature, $T_{\mathrm K}$. Thus $T_{\mathrm s}$
is a function of three local quantities:  the overdensity $\delta$, the
gas kinetic temperature $T_{\mathrm K}$ and the local Lyman-$\alpha$ flux $J_\alpha$ \citep[for details, see e.g.][]{Vonlanthen11}. Consequently, to be able to produce the 
$21$~cm data cubes, we need the local values of the overdensity $\delta$, the ionization fraction $x_{\mathrm H}$, the velocity field $v$, the temperature $T_{\mathrm K}$ and the Lyman-$\alpha$ flux $J_\alpha$. To compute these, we run a suite of three simulations: first we compute the hydrodynamics, then the ionizing UV and X-ray
radiative transfer, and finally Lyman band radiative transfer. Lyman radiative transfer can safely be decoupled from the other two and run as a post-processing: the heating of the gas by Lyman-$\alpha$ photons  is negligible compared to other sources of heating \citep[see][for quantitative evaluation]{Furlanetto06}.  The  backreaction of ionizing UV on the dynamics is effective only in haloes that have a virial temperature lower
than $10^4$ K, i.e. haloes with masses lower than $10^8 M_\odot$. In our large-volume simulation we do not resolve these haloes so we run
the ionizing UV transfer after running the hydrodynamical simulation. We now describe each step in the simulation suite.

\subsection{Hydrodynamic run}
We run the hydrodynamical
simulation using GADGET2 (\citet{Springel05}), with $2\times512^3$ particles in a
400$h^{-1}$ cMpc box.  Snapshots are extracted   using a fixed interval of the scale
factor $\Delta a=0.001$. This produces more than 100 snapshots by $z=6$, with a varying time interval of the order of $10$ Myr between snapshots. The cosmological parameters are chosen from
WMAP7+BAO+$\rm{H}_0$ data:
$\Omega_{\mathrm m}=0.272$, $\Omega_{\Lambda}=0.728$, $\Omega_{\mathrm b}=0.0455$, $h=0.704$ (\citet{Komatsu11}). 

The mass of a dark matter particle in our simulation suite is $\sim 4\times10^{10}$ M$_\odot$, and that of a baryonic particle $\sim 8\times 10^{9}$ M$_\odot$. Consequently, only haloes with masses above $\sim 5\times 10^{11}$ M$_\odot$ are resolved. It is believed, however, that 80 \% of the photons 
contributing to reionization are emitted by galaxies with masses smaller than $10^{9}$ M$_\odot$ \citep[e.g.][]{Choudhury07}. This implies that in our 
simulations, lacking the contribution of small galaxies, the luminosity of massive galaxies is boosted to complete reionization by $z=6$. This relocation 
of ionizing photons from small to massive galaxies changes the topology of the reionization field to some degree. Whether it affects the level of 
anisotropy in the $21$ cm signal is difficult to assess, but is indeed a possibility. 
A definitive answer to this question would require
running much larger simulations ($8192^3$ particles in the 
same volume), both for the dynamics and for the radiative transfer.

\subsection{ Ionizing radiative transfer run}
We next compute the radiative transfer of UV ionizing photons using the code LICORICE
\citep{Baek09, Iliev09}. In the version of LICORICE used in \citet{Baek10}  a
finite velocity ($c=3\times 10^8\rm{\,m\,s^{-1}}$) was used only for X-ray photons. UV photon packets were propagated at infinite speed until 
99.99\%  of their content was absorbed. In the version we use for this work, we implement the actual finite speed of light
for both X-ray and UV photons. Between two snapshots, photon packets typically travel 25 cMpc. During the overlap phase, ionized regions
can extend to more than 100 comoving Mpc, which are fully sampled in our 400~$h^{-1}$ cMpc  simulation volume. Since the luminosity of
the sources (updated with every new snapshot) changes substantially over a few tens of Myr, using an infinite speed of light results in overestimating
the speed of the ionization fronts. This issue is most sensitive if the simulation volume is large enough to produce very large ionized patches, which is our case. 
Another issue occurs
when dealing with large ionized patches. The estimated mean free path of ionizing photons through
the forest of Lyman Limit Systems (LLS) at $z \sim 6$ is in the $50$ cMpc range 
\citep[e.g.][]{Songaila10}. Such systems are definitely not resolved in large volume simulations 
of the EoR. Consequently, the flux received by the ionization fronts of very large patches is 
somewhat overestimated. Simulations taking LLS into account through sub-grid recipes do not suffer as much from using an infinite speed of light.We do not include any treatment of Lyman Limit Systems.

The star formation efficiency and escape fraction ($f_\mathrm{esc}=6\%$) are calibrated to reach complete reionization at $z=6$. Sources are formed in regions where the gas is above a fixed density threshold using a standard Schmidt law. We use a simple model to account for the production of X-rays by quasars, supernovae (SN)
and other possible sources: each source produces 0.6\% of its luminosity in the form of X-rays. We use a 100\% escape fraction for X-rays. 
This level of X-ray emission is equivalent to $f_{\mathrm X} \sim 5$ \citep[as defined by][ for example]{Furlanetto06}, considering that we use the same source model as in \citet{Baek10}. The level of X-ray emission quoted in \citet{Baek10}  should be corrected by $f_{\mathrm{esc}}$, since this factor was (incorrectly) applied to X-ray emission.

\subsection{Lyman band radiative transfer}

In order to compute the Lyman band radiative transfer part of the simulation pipeline, we first interpolate the output 
of the ionizing radiative transfer run on a $512^3$ grid. We then emit $1.6 \times 10^9$ photons between each pair of snapshots, and propagate them at the speed of light. The
full radiative transfer in the first five Lyman resonances is computed, assuming a flat spectrum in that range, between $z = 13.8$ and 
$z = 7.5$.  A complete description of the numerical scheme and the physics included in our simulations can be found in \citet{Semelin07}
and \citet{Vonlanthen11}. At lower redshifts, we make the assumption that the spin temperature is fully coupled to the gas kinetic temperature in the neutral regions: $T_{\mathrm s}=T_{\mathrm K}$.

\subsection{21 cm differential brightness temperature data cubes}

To investigate the correlation function and the evolution of the power spectrum we analyse the results of the numerical simulation suite, 
which consists of 77 cubes of differential brightness  temperature $\delta T_{\mathrm b}$ in the range of redshifts from $z=13.84$ to $z=6.06$. 
We compute these cubes at different resolutions for the following cases:
\begin{itemize}
              {
\item basic $\delta T_{\mathrm b}$--- excluding velocity gradients ($\delta T_{\mathrm b}$) 
\item as for basic $\delta T_{\mathrm b}$, but  taking into account the velocity gradient term ($\delta T_{\mathrm b}$+velocity gradient) 
\item basic $\delta T_{\mathrm b}$ degraded using SKA-like noise and resolution ($\delta T_{\mathrm b}$+SKA)
\item as for basic $\delta T_{\mathrm b}$, but including both velocity gradients and SKA-like noise and resolution ($\delta T_{\mathrm b}$+velocity gradient+SKA) 
\item basic $\delta T_{\mathrm b}$ degraded using LOFAR-like noise and a resolution of 15 arcmin ($\delta T_{\mathrm b}$+LOFAR 15')

}
\end{itemize}

\subsubsection{Including the velocity gradient contribution}

Local peculiar velocity gradients modify the intensity of the emitted $21$~cm signal \citep{Bharadwaj04, Barkana05}. Since their 
contribution is anisotropic, we must include them to check that they can be disentangled from the light cone effect. While the velocity
gradient term in eq. (\ref{equdtb}) is straightforward to implement, it is a linear approximation for small gradients: if
the value of a negative gradient reaches that of the local Hubble flow, the brightness temperature (unphysically) diverges. \citet{Mao12}
have proposed an improved quasi-linear scheme to compute this contribution in a way that avoids divergences. In this work we use the usual linear
approximation and implement a cut-off to avoid divergences.

Indeed, without a cut-off, the minimum and maximum values of $\delta T_{\mathrm b}$ are much below $-1000$~K and above  4000~K, respectively. These rare spurious cells strongly influence the shape of the power spectrum and correlation function, and significantly distort them.

To avoid this problem we use a clipping method, i.e. we replace cells with $\delta T_{\mathrm b}$ greater than 100~mK or smaller than $-300$~mK by cells with $\delta T_{\mathrm b}$ exactly equal to 100~mK and $-300$~mK, respectively. We estimate the robustness of this method on data containing the basic $\delta T_{\mathrm b}$ (without a velocity gradient). We check that the power spectrum and correlation function for the basic $\delta T_{\mathrm b}$ remains unchanged when data are clipped  in this range. We use a rather low cut-off since we want to be able to check that we do not alter the properties of the signal itself by using a cut-off, independently of the spurious divergent values. We also check that a much higher cut-off on data with velocity gradient works just as well to  remove the effect 
of spurious cells, and that the result is independent from the value of the cut-off, as long as we cut all cells with $\delta T_{\mathrm b}$ below $\sim -10^5$~mK and above a few $10^4$~mK.

\subsubsection{Including SKA noise and resolution}
In order to include SKA-like noise and resolution, we proceed in three steps. We first build cubes of pure noise, using the noise power spectrum given in \citet{Santos11}. 
The resolution of these noise cubes corresponds to the expected SKA resolution at the corresponding redshift. We assume a 5~km core, a total observation time of 1000~h, 
and a maximal baseline of 10 km. In order to calibrate the noise rms of our realizations, we consider the value of 1 mK at $z = 6$ mentioned in the SKA Design Reference Mission (DRM), with  1 MHz frequency resolution and a 1.1 arcmin angular resolution (although \citet{Mellema13} estimate that this level should be reached only at 5 arcmin resolution). Assuming Gaussian statistics, we normalize that value to the size of our simulation cells to get an rms of about 9 mK at $z = 6$. We then deduce the necessary sensitivity to reach this rms, and assume that the effective collecting area $A_{\mathrm{eff}}$ evolves as the square of the observed wavelength for observed frequencies above 100 MHz. For lower observed frequencies, $A_{\mathrm{eff}}$ is constant and equal to its value at 100 MHz.

Then we add the noise cubes to the full resolution $\delta T_{\mathrm b}$ cubes. We smooth the resulting data by applying a 3D Gaussian smoothing with the full width at half maximum (FWHM) equal to the resolution at the corresponding redshift.

\subsubsection{Including LOFAR noise and resolution}

To simulate LOFAR noise, we go through the same three steps as for the SKA
case. Here, we normalize the noise rms to  56 mK at 150~MHz,
as given in \citet{Zaroubi12} for an angular resolution of 3~arcmin and a
frequency resolution of 1~MHz. Following \citet{Zaroubi12}, we assume a
constant resolution of 15~arcmin at all redshifts. We scale the rms value at 15~arcmin resolution from the value at 3~arcmin by assuming that the noise is Gaussian, although this is not strictly the case. The behaviour of the rms with wavelength is illustrated in \citet{Zaroubi12}, Fig. 3, based on a detailed model of the instrument.  As a rough fit we scale the rms as  $\lambda^2$, where $\lambda$ is the observed wavelength.
A detailed model for the evaluation of the rms can be found in \citet{Labropoulos09}. We add the noise cubes to the full resolution data and
smooth the result to the expected LOFAR resolution (15~arcmin).

\subsection{Building light cones}

\begin{figure}
\includegraphics[width=84mm]{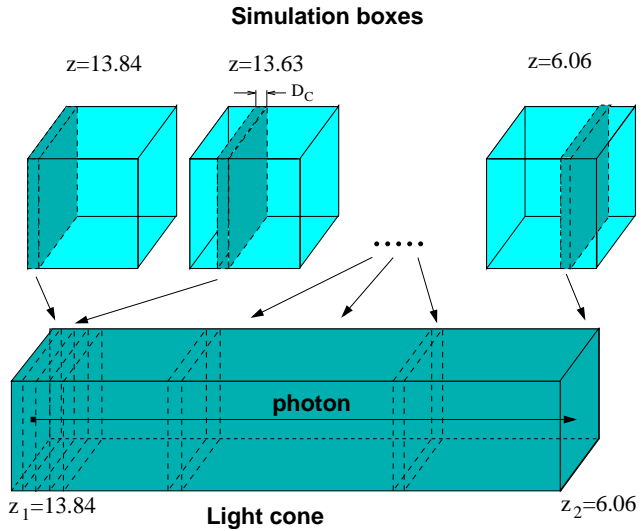}
 \caption{The light cone which models the Universe.}
  \label{fig:lc}
  \end{figure}  
First, we make a flat sky approximation. This introduces a distortion
of $\sim 4$~cMpc along the LOS at the edge of a $3^\circ$ field of view at
$z=13.6$. This is
close to the SKA resolution, which is acceptable for the purposes of this paper. Second,
although we will be refer to a light cone, we actually build a light cylinder, assuming that all the lines of sight are parallel. On the scales where the anisotropy
is found ($100$~cMpc), a real light cone would involve $\sim 1\%$ change in 
the linear size for two cross-sections $100$~cMpc apart along the LOS. The effect would be negligible on the correlation function amplitude. One non-negligible effect would be to increase the size of the sample of pairs of points between $z=6$ and $z=13.6$ by up to a factor of two, and thus possibly induce a slight change in the variance of the correlation function. In view of other approximations made in our method, we feel that the light cylinder approximation is acceptable.

When building a light cone from simulations, a common trick is to translate and rotate the box every time the light cone enters a new periodical replica of the simulation box. This avoids encountering the same structure repeatedly along the line of sight and other artefacts of using a simulation box that is smaller than the length of the light cone.
While this works well
in the context of  building galaxy catalogues, which are discrete samples, it cannot be applied to $21$~cm light cones. Indeed, large ionized regions straddling the box boundary would be truncated 
without any possible justification. Consequently, we must forgo these rotations and translations.
To avoid the artefacts, we must use a large simulation box. With our $400\, h^{-1}$~cMpc box, we need only
three replications along the light cone.

The light cone is made of a series of slices, each cut from a different snapshot (Fig. \ref{fig:lc}). The thickness of the slices that form the light cone is
\begin{eqnarray}
D_{\mathrm C} = \frac{c}{H_0} \int_{z_{\mathrm A}}^{z_{\mathrm B}} \frac{{\mathrm d}z}{\sqrt{\Omega_M (1+z)^3 +\Omega_\Lambda}},
\label{equDc}
\end{eqnarray}
where $z_{\mathrm A}$ and $z_{\mathrm B}$ are the redshifts of two consecutive snapshots.
The collated slices constitute the light cone from $z=13.84$ to $z=6.06$. The light cone has a size of 400 $h^{-1}$ cMpc $\times$ 400 $h^{-1}$ cMpc $\times$ 1.97~cGpc (512 $\times$ 512 $\times$ 1774 cells), so one cell corresponds to 1.1~cMpc.

To improve statistical significance we build separate light cones along the X, Y, and Z axes of the simulation boxes. We also rotate the simulation 
boxes 45 degrees around the X, Y, or Z axis and build the light cones along the Y and Z, or Z and X, or X and Y axes, respectively. 
 In this way we generate 9 quasi-independent axes of space-time. Along each of these 9 axes we build four light cones, starting from different positions in the first snapshot, namely from cell index 1, 128, 256 and 384.
We use these 4 light cones generated along the same axis to check whether the periodicity conditions affect the outcome of our analysis. If periodicity affected the correlation functions, it would create characteristic features in the curves that would be simply differently shifted between the 4 cases. We do not observe any such effect. Altogether, we create 36 quasi-independent light cones. 

\begin{figure}         
 \includegraphics[width=1.0\linewidth]{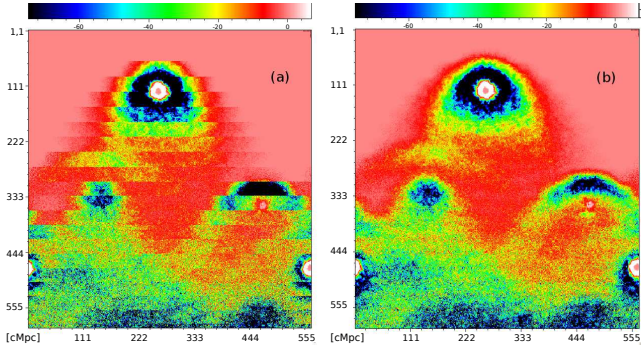} 
 \caption{(Colour online) A fragment of the cross section of the light cone (512x512 cells) along the space-time axis  before (a) and after interpolation (b). False colours describe the differential brightness temperature $\delta T_{\mathrm b}[mK].$ Units on the $x$ and $y$ axes are in cMpc.}
 \label{fig:lc2}
 \end{figure}

The light cone at low redshift is quite smooth, while at high redshift a banding is visible due to the slice structure (Fig. \ref{fig:lc2}(a)). Extracting the snapshots with shorter time intervals is possible but costly in terms of CPU and storage. 
To eliminate these discontinuities we linearly interpolate between $\delta T_{\mathrm b}(\vec{x},z_i)$ and $\delta T_{\mathrm b}(\vec{x},z_{i+1})$ as a function of redshift, between consecutive snapshots $i$ and $i+1$ at redshifts $z_i$ and $z_{i+1}$ (Fig. \ref{fig:lc2}(b)).

The interpolation smooths the structure at high redshift and does not  distort it at low redshift. Thus, all our calculations are based on the interpolated light cones. Examples of the cross section of different light cones are depicted in Fig.~\ref{bbSKA}.

\subsection{Correlations}

Two points separated by a distance $r$ from each other and observed along the LOS are seen at different redshifts. The time-varying distribution of the HII regions
 influences the correlation function, which is an average over such pairs of points. We follow the definition of the correlation function $\xi$ formulated by \citet{Barkana06} to examine the light cone  anisotropy. In this model the two-point correlation function is expressed by the equation
 \begin{eqnarray}
\xi(\delta T_{\mathrm b},\Theta,r,z) = \left\langle \left[ \delta T_{{\mathrm b},1} - \delta \bar{T}_{\mathrm b}(z_1)\right]  \left[ \delta T_{{\mathrm b},2} - \delta \bar{T}_{\mathrm b}(z_2)\right] \right\rangle,
\end{eqnarray}
where
\begin{itemize}
              {
              \item $\xi$ is parametrized as a function of  $\Theta$, the angle between the LOS and the line connecting the two points. $\Theta = 0^\circ$ means 2 points along the LOS and $\Theta = 90^\circ$  means 2 points whose separation vector is perpendicular to the LOS,
              \item $\xi$ is a function of the comoving distance $r$ between two points at 
two different redshifts ($\Theta = 0^\circ$) or at the same redshift ($\Theta = 90^\circ$). We calculate  $\xi$  for three distances: $r$=100~cMpc, 50~cMpc, 20~cMpc,

\item in the $\Theta = 0^\circ$ case, a single redshift $z$ is taken at the midpoint between the pair of points 
in terms of comoving distance,
\item the mean $\delta \bar{T}_{\mathrm b}$ at the given redshifts is subtracted from $\delta T_{\mathrm b}$ at each cell because this is what upcoming interferometer data will look like: they will only contain  fluctuations of $\delta T_{\mathrm b}$.

                    } 
              \end{itemize}
             
Since variance will prove to be an issue, the choice of binning will be critical when
presenting our results for the correlation function. Our resolution element is a cubic 
cell $1.1$ cMpc in size. Thus, naturally, our bin size for $r$ is $\Delta r=1.1$~cMpc. The
bin size for $\Theta$ is then $\Delta\Theta=0.63^\circ$ for $r=100$~cMpc,  $\Delta\Theta=1.26^\circ$ for $r=50$~cMpc and  $\Delta\Theta=3.15^\circ$ for $r=20$~cMpc. Using a single cell for the
bin size in $z$ would yield very noisy results. Instead we use bins of 40 cells, which correspond to
$\Delta z = 0.29$ at $z=13.8$ and $\Delta z= 0.1$ at $z=6.$ These bins overlap since
we use steps of 5 cells between them. A much larger $\Delta z$ bin size would smooth the
features in the evolution as a function of $z$ and thus weaken the discriminating power of
$\xi$. Using larger bin sizes for $r$ and $\Theta$ would
not probe information independent of that found using the large $\Delta z$ bin size. Finally, we
average over the 36 light cones. We could have used a binning system closer to observations, using constant bin size in frequency and angular resolution instead of redshift and comoving position. However, this would
involve a lot of interpolation and would only introduce small differences in the error bars.

\section{Results}

\subsection{ Global reionization history}

\begin{figure}            
 \includegraphics[width=0.7\linewidth,angle=-90]{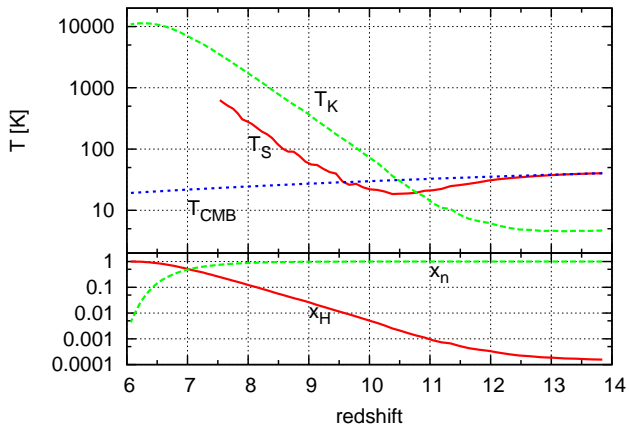}
 \caption{Top: The box-averaged spin, kinetic and CMB temperatures as a function of the redshift. At redshifts
lower than 7.5, we make the assumption that, in neutral regions, the spin temperature is fully coupled to the gas kinetic temperature: $T_{\mathrm s}=T_{\mathrm K}$.
 Bottom: Cosmic neutral fraction $x_{\mathrm n}$ and the ionization fraction $x_{\mathrm H}$ as a function of redshift  during the process of cosmic reionization.}      
 \label{Global_history}       
\end{figure}

We have calibrated the product of the escape fraction and the star formation efficiency to reach complete reionization at $z=6$, in agreement
with observations of the Gunn--Peterson effect in the spectra of high redshift QSOs \citep[e.g.][]{Fan06}. The resulting global ionization history
is shown in Fig. \ref{Global_history}, along with the evolution of the average spin and kinetic temperatures. The corresponding Thomson scattering optical depth is $\tau=0.054$. This value is  $2.5\,\sigma$ 
below the value inferred from observation of the CMB \citep[$\tau= 0.089 \pm 0.014$,][]{Planck13}. This is common
for numerical simulations of the EoR that end reionization at $z=6$:
it is difficult to match both observational constraints simultaneously. Although
a higher resolution allows starting reionization earlier, it does not
seem to be enough to bring $\tau$ in agreement with observation while completing
reionization at $z=6$. This fact suggests that either we misinterpret the observations or that something is missing in the simulations \citep[e.g.][]{Ahn12}.

\citet{Barkana06} considered two different ionization histories, early reionization  driven by population III stars (Pop III), and late reionization driven by Pop II
stars. Their early reionization scenario, however, is rather unlikely  since it completes reionization at $z \sim 13$, and is presented mainly as
a point of comparison. Our reionization history is similar to their late reionization scenario.

\subsection{ Evolution of the power spectrum }

\begin{figure}
\includegraphics[width=0.7\linewidth, angle=-90]{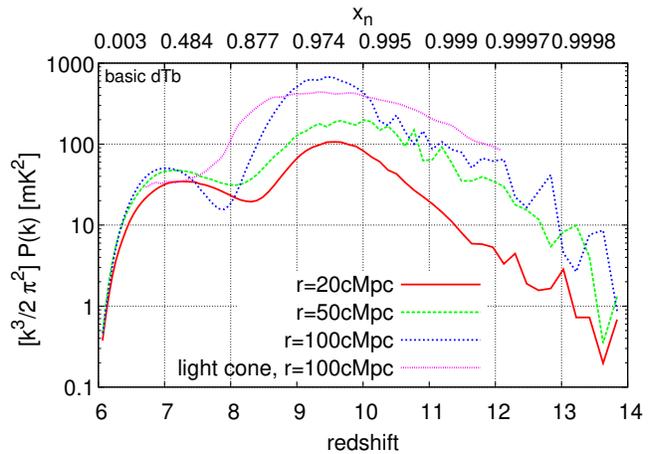}
\caption{Evolution of the differential brightness temperature power
spectrum with redshift at wavevectors which correspond to distances of 20, 50 and 100~cMpc. Magenta (dotted) curve shows the power spectrum at 100~cMpc computed from the light cone.}
\label{power_log}
\end{figure}

Figure~\ref{power_log} shows the evolution of the differential brightness temperature power spectrum as a function of redshift for three different wavevectors
corresponding to wavelengths of 20, 50 and 100~cMpc. We present these plots mainly to allow comparison with previous works. The evolution
is in agreement at a large scale with model S7 in \citet{Baek10},
which also assumes strong X-ray heating 
(their Fig.~6).
Our result is also in general agreement with \citet{Santos08}. The magenta curve in Fig.~\ref{power_log} shows the evolution of the power spectrum computed  from $400\, h^{-1}$~cMpc boxes cut from the light cone, for wavevectors which correspond to distances of 100~cMpc. The bumps are smoothed in this case  because within the boxes used to compute the power spectrum, one finds different stages in the history of reionization, and 
when $P(k)$ is computed for a given $k$ value, it is averaged over a whole era 
of reionization history. Indeed, $400\, h^{-1}$~cMpc is more than the (comoving) distance travelled by light between redshift 8 and 6, and between these two redshifts, in our simulation, 
the ionization fraction rises from 0.1 to 1. In the snapshot version, bumps have maxima at 
specific $z$. In the light-cone version, one averages over a non-negligible 
$\Delta z$, so the bumps are smoothed. We observe the smoothing effect at all scales studied. This is similar to what \citet{Datta12} found with a $163$ cMpc box size, showing that the 3D isotropic $P(k)$ is not a good diagnostic when applied to light-cone data on large scales. Consequently, and to facilitate comparison with \citet{Barkana06}'s results, we will focus our analysis on the correlation function.

\subsection{ The effect of sample variance}

\begin{figure}
 \includegraphics[width=0.7\linewidth,angle=-90]{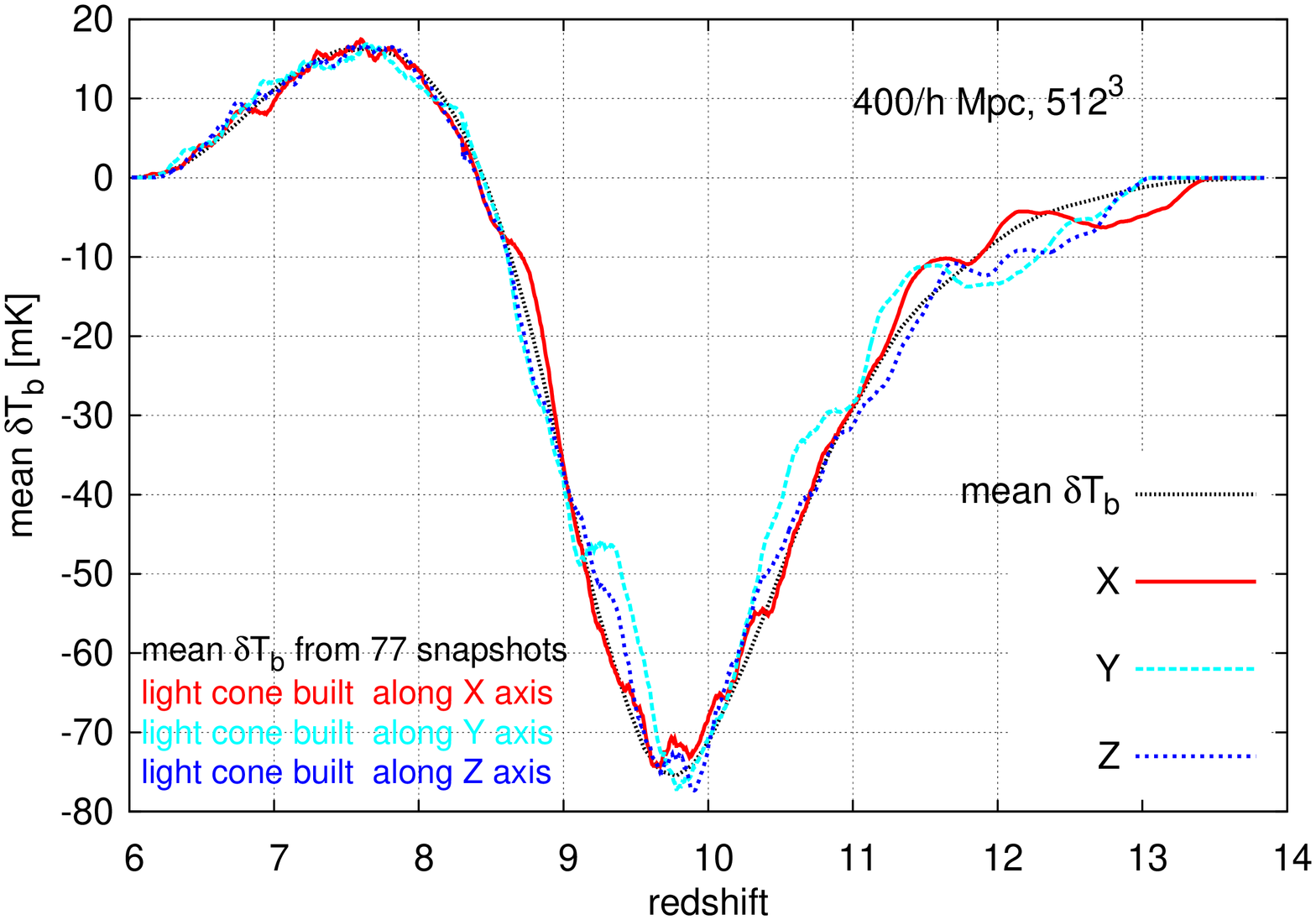}
\includegraphics[width=0.7\linewidth,angle=-90]{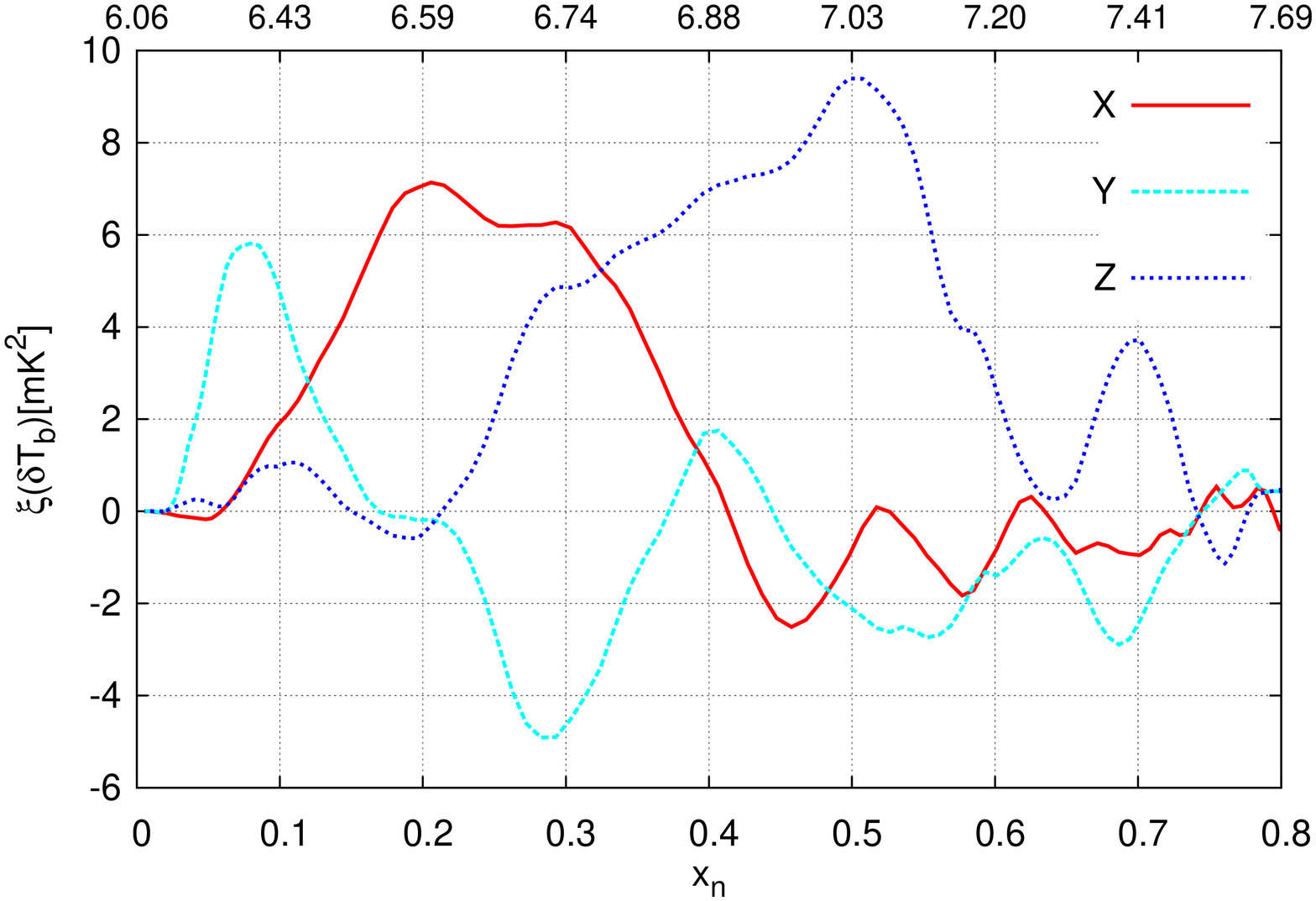}
 \caption{(Colour online) Top: The mean differential brightness temperature $\delta T_{\mathrm b}$ of the $21$~cm signal from simulation snapshots of basic $\delta T_{\mathrm b}$ (black smooth dotted line). Colour curves are the mean $\delta T_{\mathrm b}$ along the light cones built along the X axis (solid red), Y axis (dotted cyan) or Z axis (short dotted blue). Bottom: Example of the correlations functions at $r=100$~cMpc between points perpendicular to the LOS ($\Theta=90^\circ$). The calculations were made on the basis of the light cones formed along the axes X~(solid red), Y~(dotted cyan), or Z~(short dotted blue) axes. The large differences
reveal a large sample variance.}
 \label{mean_dTb_alongXYZ}
\end{figure}

Compared with the 3D isotropic correlation function, the correlation functions computed transversely, and even more so along the LOS, represent much smaller sample sizes, and thus are more noisy.

While
we could compensate for this sample-size noise by initializing the high-$z$ end of the light cone at different slices in our simulation snapshot, this will not be an option with
observational data. Consequently we need to estimate how large a survey size is needed to mitigate sample variance. 
The values that are usually quoted for the isotropic density field are of no help in our case.

Figure~\ref{mean_dTb_alongXYZ} (top panel) shows the fluctuations in the mean $\delta T_{\mathrm b}$ calculated for the light cones created along three perpendicular directions. Each
value in these curves is an average computed in  40-cell--thick slices ($\sim 30\, h^{-1}$~cMpc), perpendicular to the LOS. While this is a volume
equivalent to a $\sim$ $(170\, h^{-1}\mathrm{cMpc})^3$ cube, we still find a non-negligible variance. When we consider the evolution of the  correlation perpendicular to
the line of sight at $\sim$~100~cMpc (Fig.~\ref{mean_dTb_alongXYZ}, bottom panel), the variance is much worse. Obviously, a $3^\circ 20' \times 3^\circ 20'$ survey size
is insufficient to mitigate the variance on such directional diagnostics.

To quantitatively estimate the variance in a $3^\circ 20' \times 3^\circ 20'$ survey size, we built, as stated above, 36 nearly independent light cones. We rotated the LOS by $\pm 45^\circ$ from each of the three axes of the grid, resulting in 9 different lines of sight and used 4 different starting positions for each LOS, 
separated from one another by $100$ $h^{-1}$~cMpc in our highest redshift data cube. We estimate the sample variance for a $3^\circ 20' \times 3^\circ 20'$ survey size 
by computing the sample standard deviation for each point in the curve  over the 36 values obtained from the 36 light cones. Finally, 
for a rough estimate of the  error in a $20^\circ \times 20^\circ$ survey size, i.e. in a solid angle 36 times greater, 
we {\sl assume that this consists of 36 independent samples with Gaussian error distributions in each} (i.e. we neglect cross-correlations), and thus divide the sample standard deviation by $\sqrt{36}$.
It turns out that using the average over such a survey size allows us to differentiate the correlation along and perpendicular to the LOS at the level of $\sim 2\sigma$ at the redshifts when
the anisotropy is greatest.

\subsection{ Anisotropy in the raw signal}

\begin{figure*}
\includegraphics[width=0.5\linewidth, angle=-90]{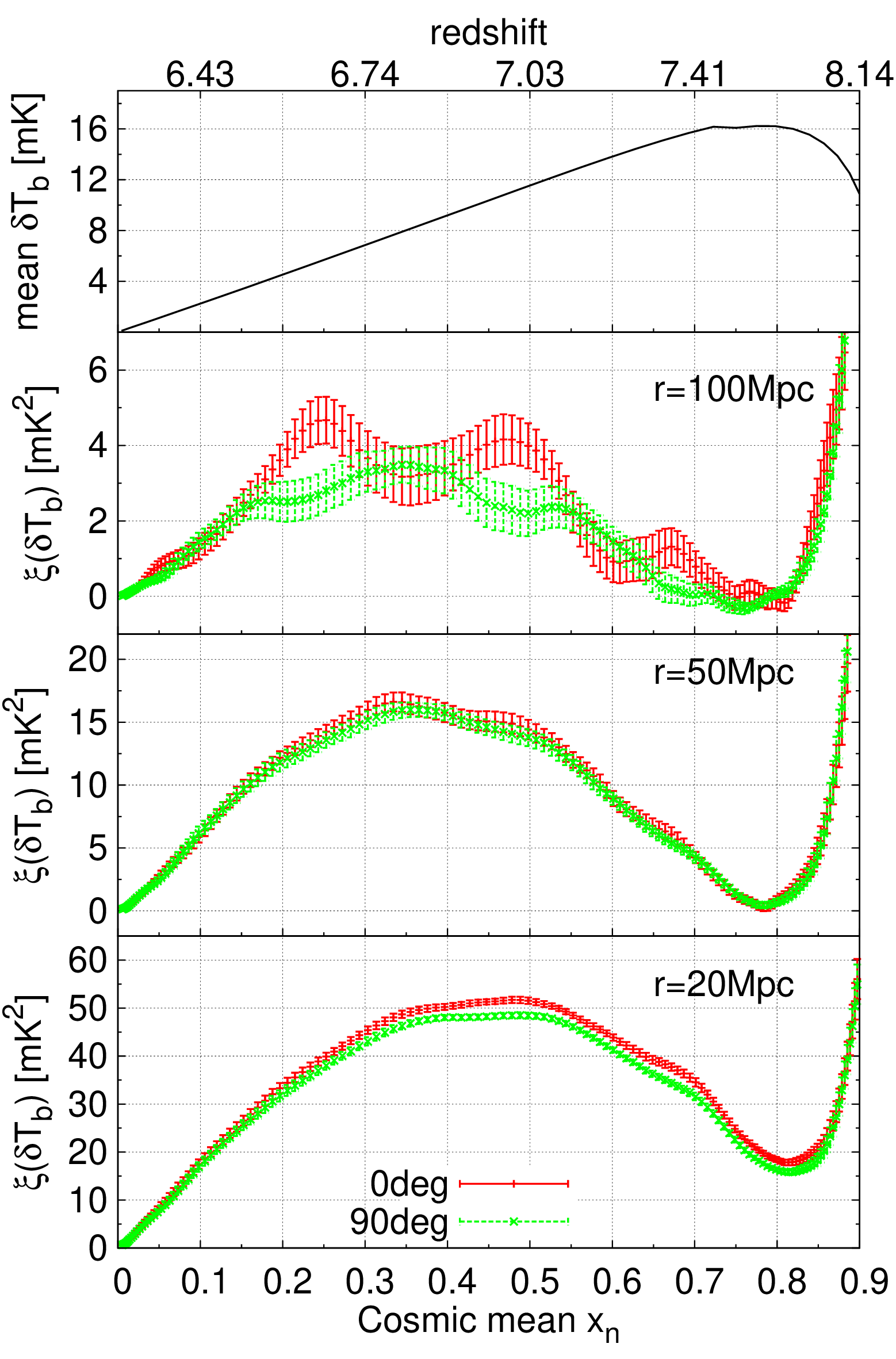}
\includegraphics[width=0.5\linewidth, angle=-90]{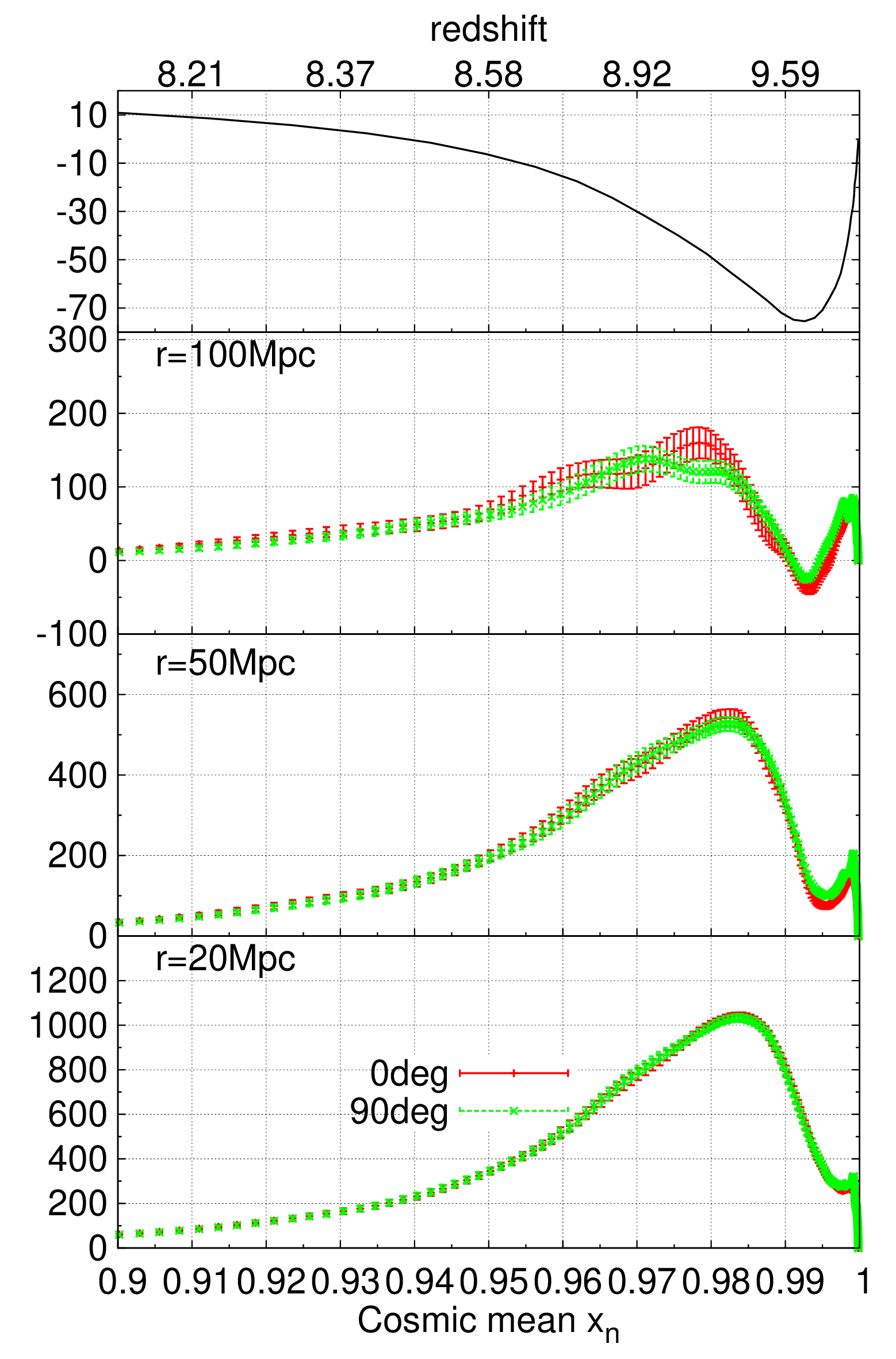}
\includegraphics[width=0.5\linewidth, angle=-90]{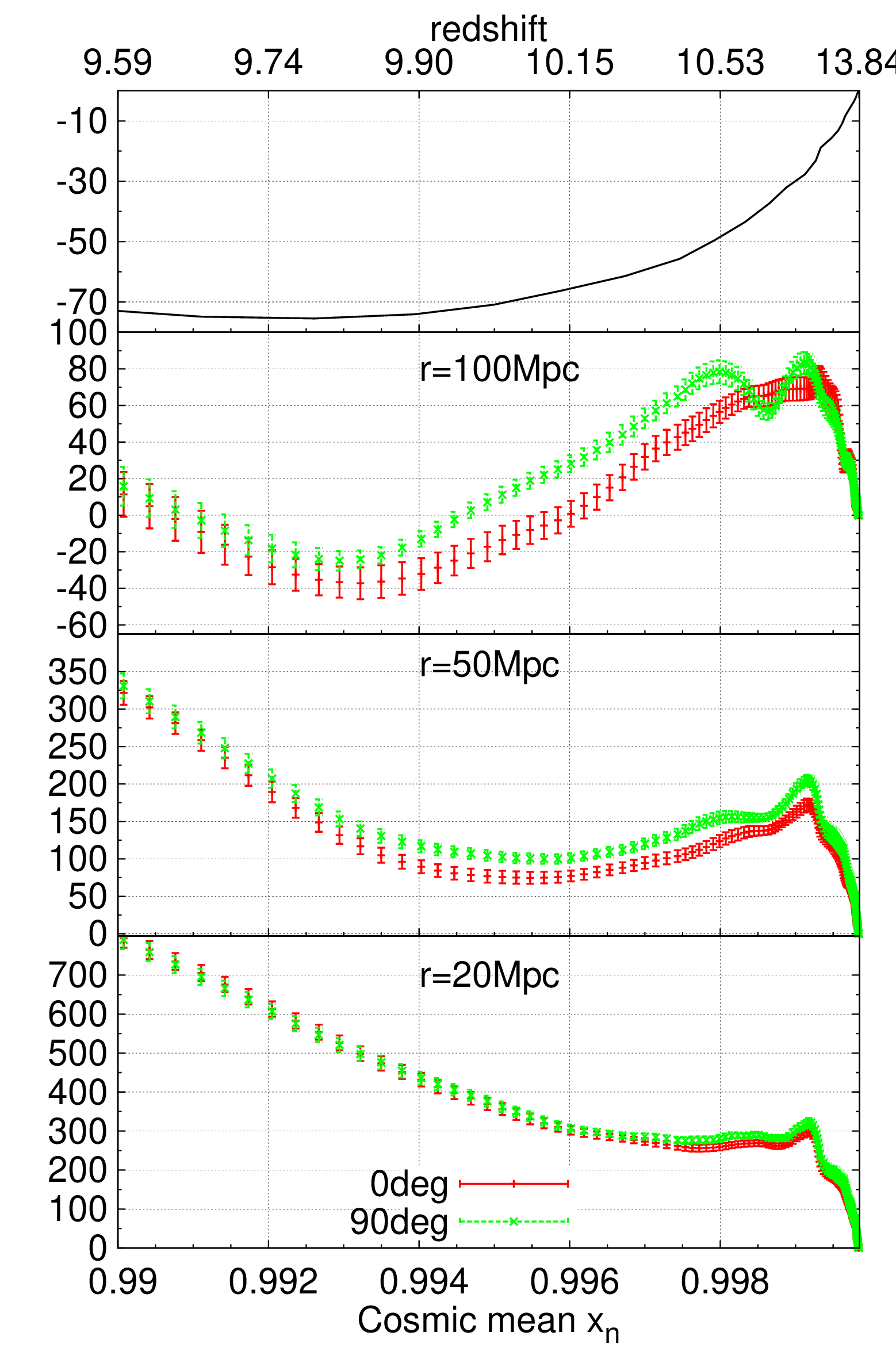}
\caption{Correlation function of $\delta T_{\mathrm b}$ simulations at different redshifts expressed as the neutral fraction $x_{\mathrm n}$ in the range (0 -- 0.9), (0.9 -- 1) and (0.99 -- 1) from left to right. Red (solid) and green (dotted) curves correspond to $0^\circ$ and $90^\circ$, respectively. The curves are averaged over 36 different light cones, and error bars are estimates of the standard deviation of the mean. The top panels show the mean $\delta T_{\mathrm b}$ from the original 77 snapshots. }
\label{corr_dTb}
\end{figure*}

Averaging over the 36 light cones, we compute the correlation function of the fluctuations in the differential brightness temperature at three different
scales: 20, 50 and 100~cMpc. The results are presented in Fig.~\ref{corr_dTb}. In these plots, the error bars include only the contribution of sample variance.

In the emission regime (left panel, $x_{\mathrm n} < 0.9$), we find correlations whose amplitudes are in general agreement with the theoretical predictions in
\citet{Barkana06} at all three scales. We also agree with them in the point that the relative amplitude of parallel to perpendicular correlations shows non-negligible anisotropy only on scales of $\sim 100$ ~cMpc.  However, BL find that perpendicular and parallel correlations peak at different stages of
the reionization history, with the correlation along the line of sight peaking first. 
We find a more complicated pattern, where the correlation along the LOS rises up to more than 2$\sigma$ above the correlation perpendicular to the LOS, in two to three different peaks. The statistical significance of this behaviour is not strong enough that we should feel entitled to attempt a detailed interpretation of those 2--3 peaks. Only the general dominance of the correlation along the LOS seems robust at this point. The behaviour found by BL disagrees with our results by more than 3$\sigma$. However, the model of BL relies on 
bubbles of fixed sizes (evolving with redshift), without accounting for the total effect occurring during the overlap phase. This is exactly when they
find an anisotropy. BL's model uses the semi-analytical method of \citet{Furlanetto04} to compute the ionization field and \citet{Santos08} found that this
method shows the largest  difference with full simulations around $x_{\mathrm n} \sim 0.2$, again where the anisotropy is found. Consequently we do not expect to find detailed quantitative agreement with BL.

\begin{figure}
\includegraphics[width=1\linewidth]{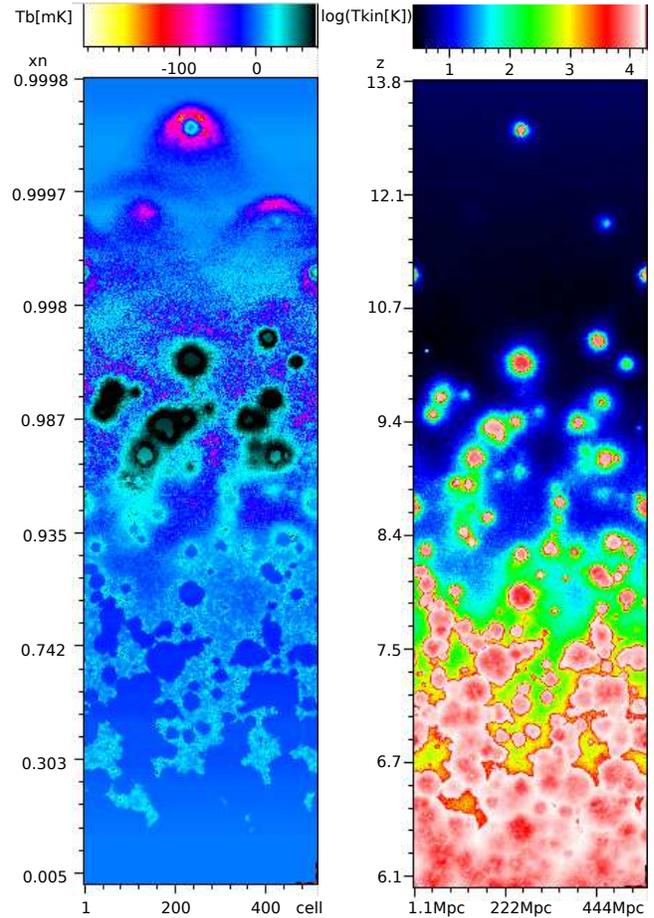}
\caption{(Colour online)  Examples of the same cross section of the light cone along a space-time axis from the basic $\delta T_{\mathrm b}$ case.
Colours on the left panel depict the difference between $\delta T_{\mathrm b}$ at given point and mean $\delta T_{\mathrm b}$ at a given redshift: $T_{\mathrm b} = \delta T_{\mathrm b} - \bar{\delta T_{\mathrm b}}$ in mK; the right panel shows kinetic temperature in K.}
\label{dTb-meandTb}
\end{figure}

\begin{figure}
\includegraphics[width=1.0\linewidth]{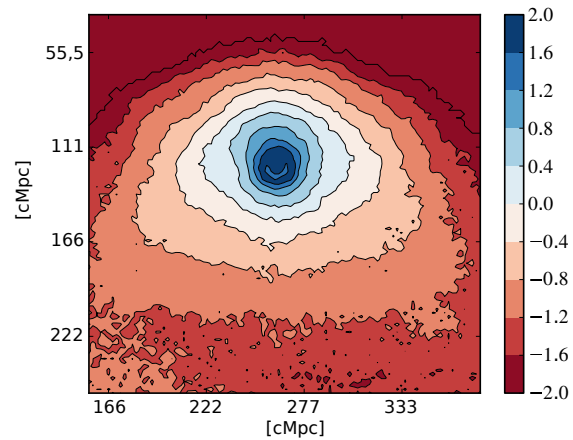}
\caption{(Colour online)  Fragment of a cross section of the light cone showing the first source. The colours depict the $x_\alpha$ factor on the logarithmic scale. Values below $10^{-2}$ are set to $10^{-2}$.
The range of the $x$ and $y$ axes (in cMpc) corresponds to a linear size of 220~cMpc. Values of cells on the $y$ axis  correspond to a range of neutral fraction $x_{\mathrm n}$ from 0.99984 to 0.99973.}
\label{x-alpha}
\end{figure}

BL's model is restricted to the emission regime when the spin temperature is assumed to be fully coupled to the kinetic temperature of the gas, itself much larger than
the CMB temperature. Our simulation suite includes 3D radiative transfer for both the inhomogeneous heating by X-rays, and the inhomogeneous coupling through Lyman-$\alpha$ photons.
So we can explore the light cone anisotropies in the absorption regime during the early EoR. In the middle panel of Fig. \ref{corr_dTb}, the 100~cMpc correlation along the
LOS rises about 2~$\sigma$ in terms of sample  variance above the correlations perpendicular to the line of sight. This happens when the Universe is $\sim
2\%$ ionized, as the average differential brightness temperature rises again after the minimum in $\delta T_{\mathrm b}$, i.e. the maximum  in absorption. While the anisotropy in the emission regime, at $x_{\mathrm n} \sim 0.2 - 0.5$, was created
by the time evolution of the ionization field, it is likely that the anisotropy at $x_{\mathrm n} \sim 0.98$ is created by the time evolution of the kinetic temperature fluctuations.
Figure~\ref{dTb-meandTb} qualitatively confirms this interpretation. It shows a crosscut of the light cone for $\delta T_{\mathrm b} - \delta \bar{T}_{\mathrm b}$ and $T_{\mathrm K}$. Structures at $x_{\mathrm n} \sim 0.2$ -- $0.5$
and $x_{\mathrm n} \sim 0.98$ seems to be more extended along the line of sight: growing bubbles of ionization ( $x_{\mathrm n} \sim 0.2$ -- $0.5$) and heated IGM regions ($x_{\mathrm n} \sim 0.98$) overlap preferentially along the time axis. 
In the right panel of Fig. \ref{corr_dTb}, anisotropy is again visible at the 2 to 3$\sigma$ level in terms of sample variance at $100$ cMpc scales, but also, to a lesser degree, in 50 cMpc correlations. It occurs in the $x_{\mathrm n}=0.994$ -- $0.999$ range and is probably created by the time evolution of
coupling fluctuations: growing regions of coupled IGM. As we can see in Fig. \ref{x-alpha}, where the isocontours of the coupling coefficient $x_\alpha$ in a region around one of the first isolated sources are shown, the coupling fluctuations are clearly anisotropic, revealing a parabolic envelope on large scales. The axis of the paraboloid is the axis of the light cone. The analytic formula for this parabola is very simple to establish if one assumes a source of constant luminosity switching on at a given time, and that all the lines of sight in the region are parallel to each other (small angular size).

\subsection{ Effect of peculiar velocities}

\begin{figure*}
\vbox to 142mm {
 \includegraphics[width=0.73\linewidth,angle=-90]{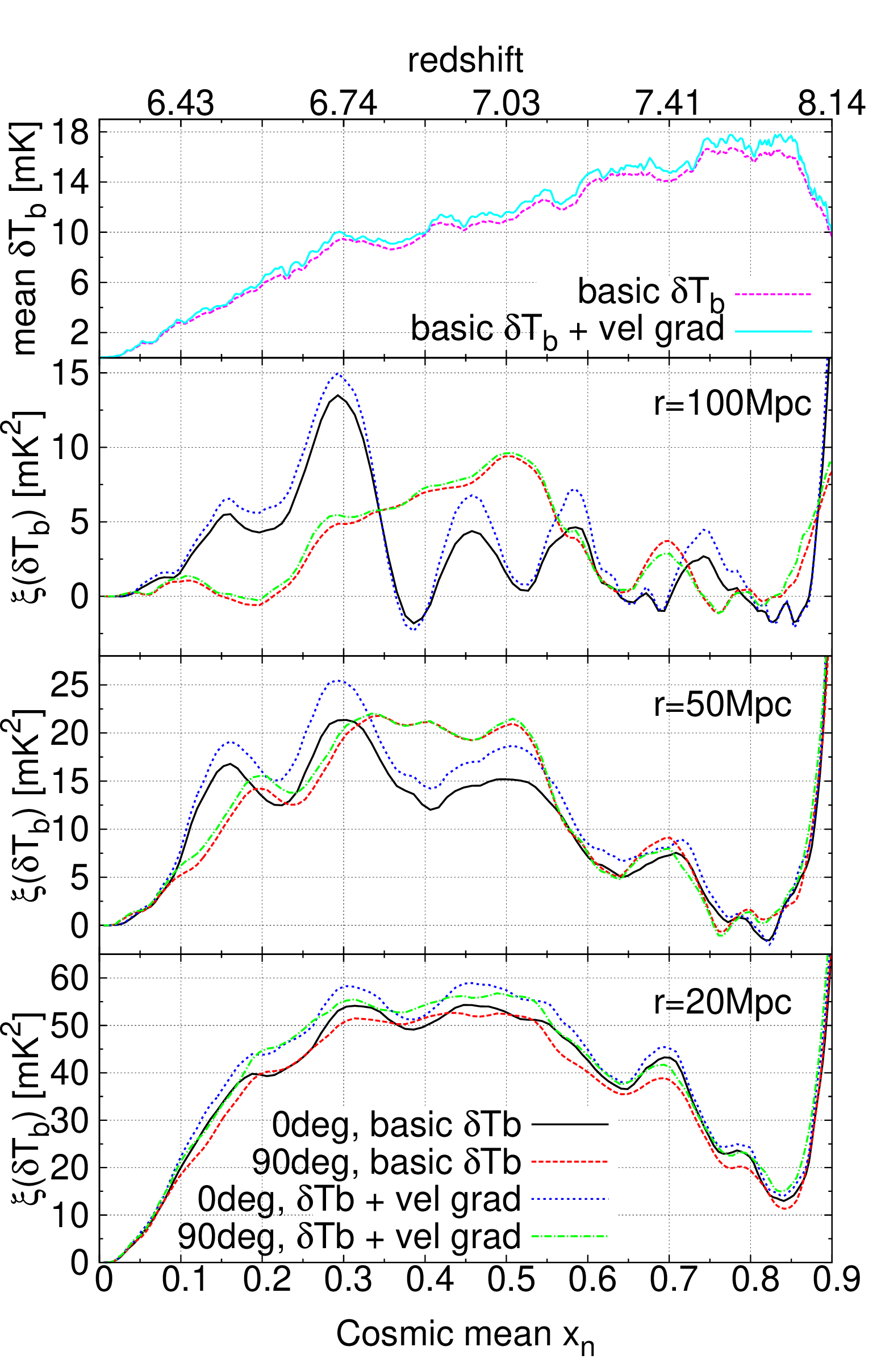}
 \includegraphics[width=0.73\linewidth,angle=-90]{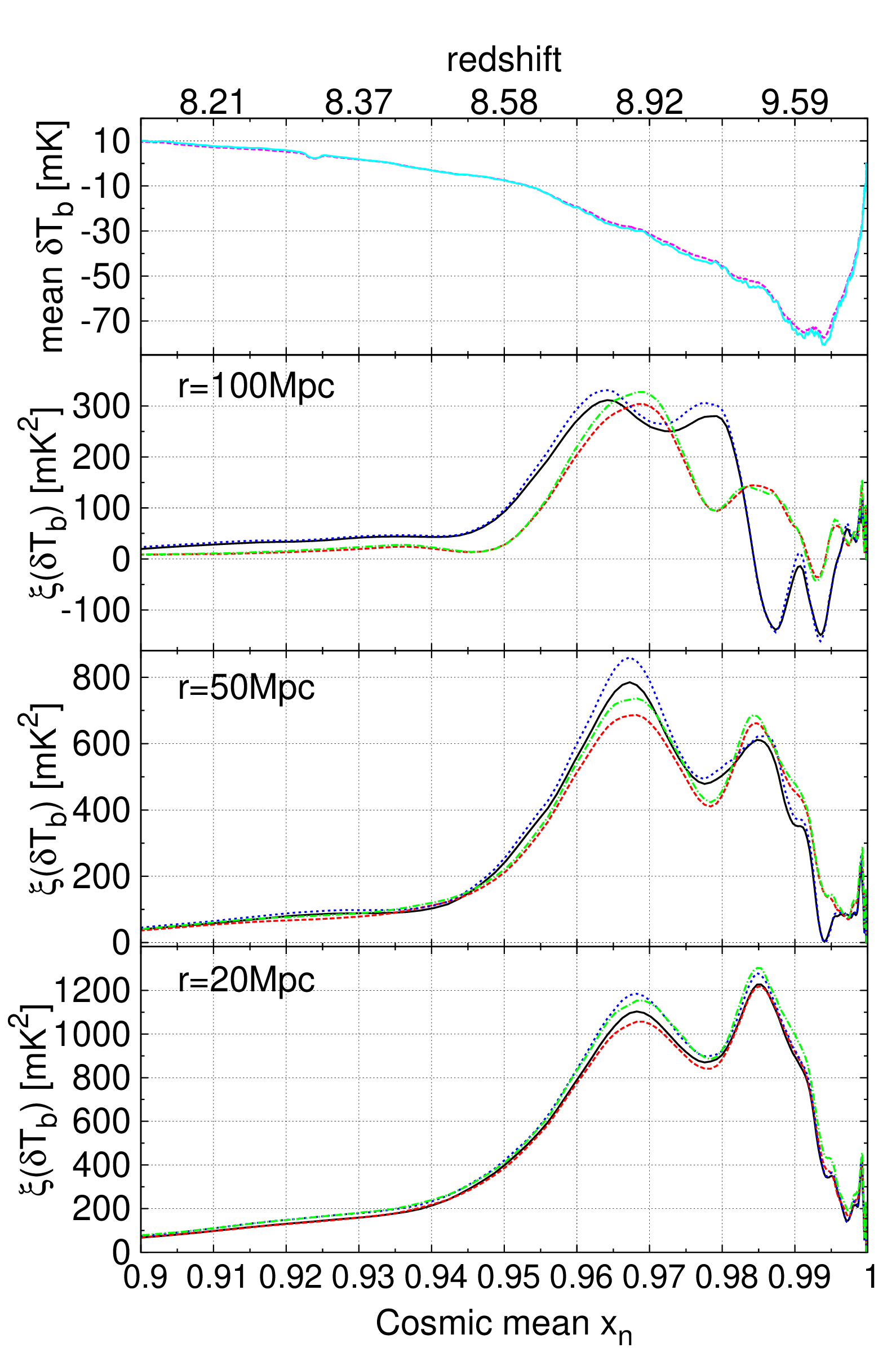}
 \caption{(Colour online) Example of the correlation functions obtained from a single light cone. Correlations along ($0^\circ$) and perpendicular ($90^\circ$) to the LOS are presented for two cases: basic $\delta T_{\mathrm b}$ and $\delta T_{\mathrm b}$+velocity gradient. The left and right panels show different ranges for the neutral fraction $x_{\mathrm n}$. Peculiar velocity increases the amplitude of the correlation by a moderate amount but does not significantly change  the overall shape of the curves.}
\label{different_axes}
}
\end{figure*}

Peculiar velocity gradients are a source of anisotropic fluctuations of $\delta T_{\mathrm b}$ \citep{Barkana05}. Thus, we have to determine whether, on 100~cMpc scales,
they can be disentangled from the light cone anisotropy. In Fig.~\ref{different_axes} we compare the correlation functions of $\delta T_b$ with and without
the contribution from peculiar velocity gradients, computed from a single light cone so as not to fold in variance effects. The velocity gradient term moderately
increases the amplitude of the correlation but does not significantly change  the overall shape of the curves. On the $100$ cMpc scale, the net effect is generally smaller than the  variance
for a $20^\circ \times 20^\circ$ light cone. We conclude that on the 100~cMpc scale, when light cone anisotropy peaks, it dominates over peculiar velocity anisotropy. However, we do apply an approximate treatment of the effect of peculiar velocities, which is valid in the linear regime. A definitive answer would
require a treatment such as that advocated by \citet{Mao12} or \citet{Jensen13}.

\subsection{ Anisotropies observed with LOFAR and SKA}

\begin{figure*}
{\includegraphics[width=1\linewidth]{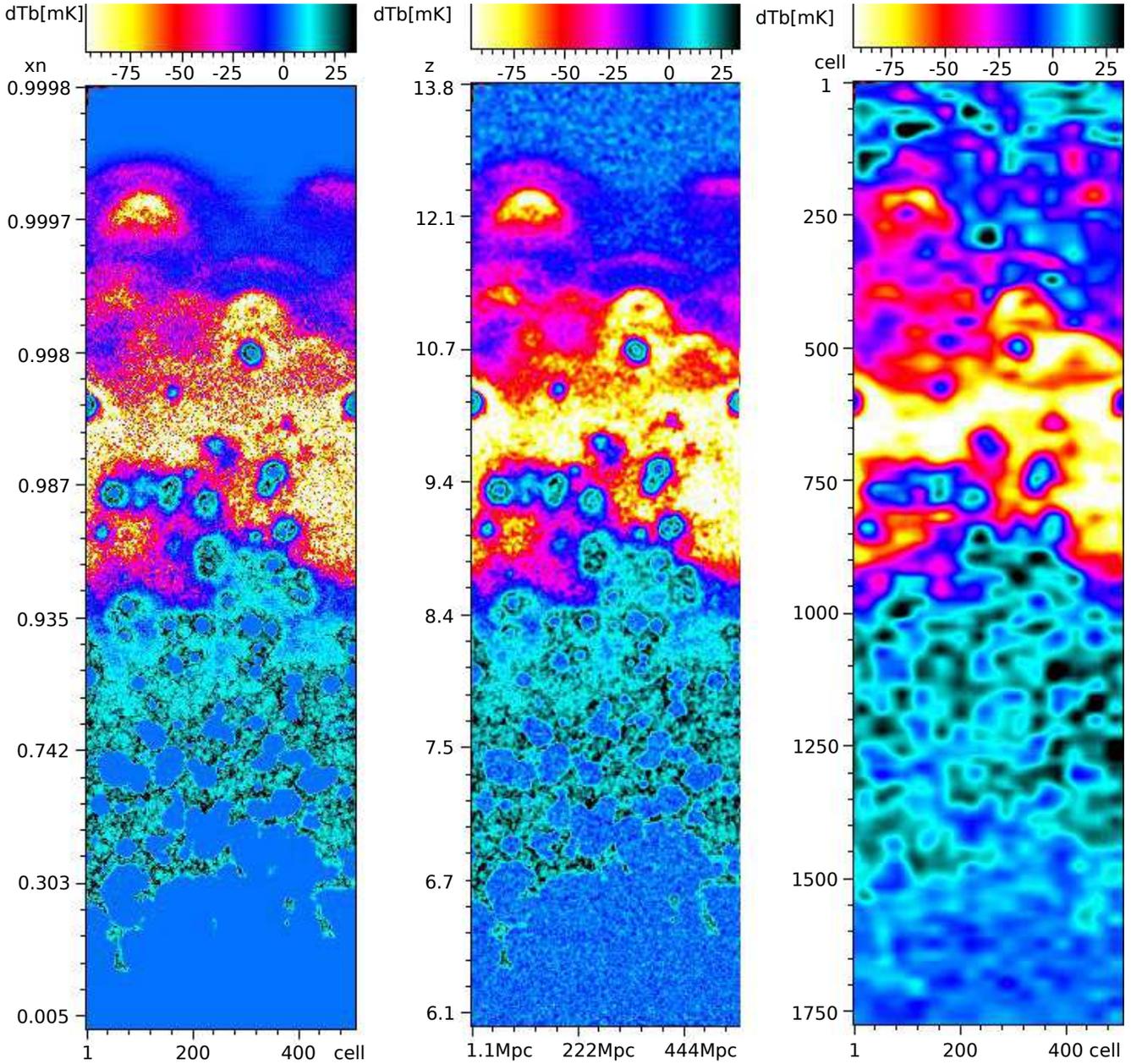}}
 \caption{(Colour online) Example of cross sections of the light cones: basic $\delta T_{\mathrm b}$ (left), $\delta T_{\mathrm b}$ with SKA-like noise simulation (middle) and $\delta T_{\mathrm b}$ with LOFAR-like noise (right). False colours show the differential brightness temperature $\delta T_{\mathrm b}[mK]$. For better visualization, 5\% of the brightest and 5\% of the darkest pixels are not shown. The $y$ axes in three pictures are the space-time axes and are denoted in three mutually dependent units: the neutral fraction of hydrogen $x_{\mathrm n}$, redshift and cells.}
 \label{bbSKA}
\end{figure*}

Will it be possible to detect the light cone anisotropy in LOFAR and SKA observations? We 
make the optimistic assumption that foregrounds have been subtracted with negligible residuals. We model only two instrumental effects: the noise
resulting from limited sensitivity and integration time, and the limited resolution (see Sect. 2.4). Figure \ref{bbSKA} shows how the cosmological $\delta T_{\mathrm b}$ light cone is modified when observed with SKA at resolution $\sim 1.5$ arcmin (depending on redshift) and with LOFAR at resolution $15$ arcmin. 

\begin{figure*}
{
\includegraphics[width=0.6\linewidth, angle=-90]{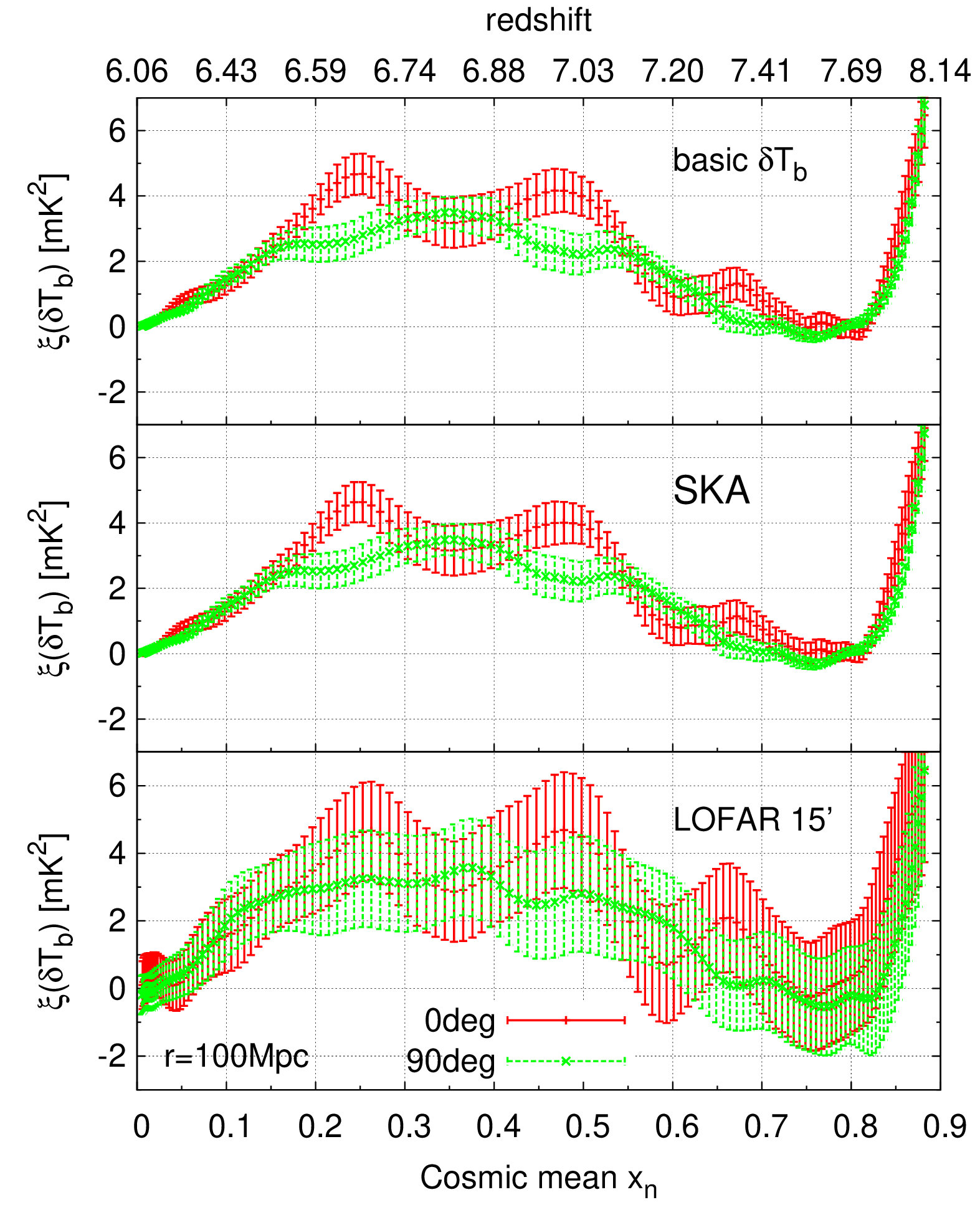}
\includegraphics[width=0.6\linewidth, angle=-90]{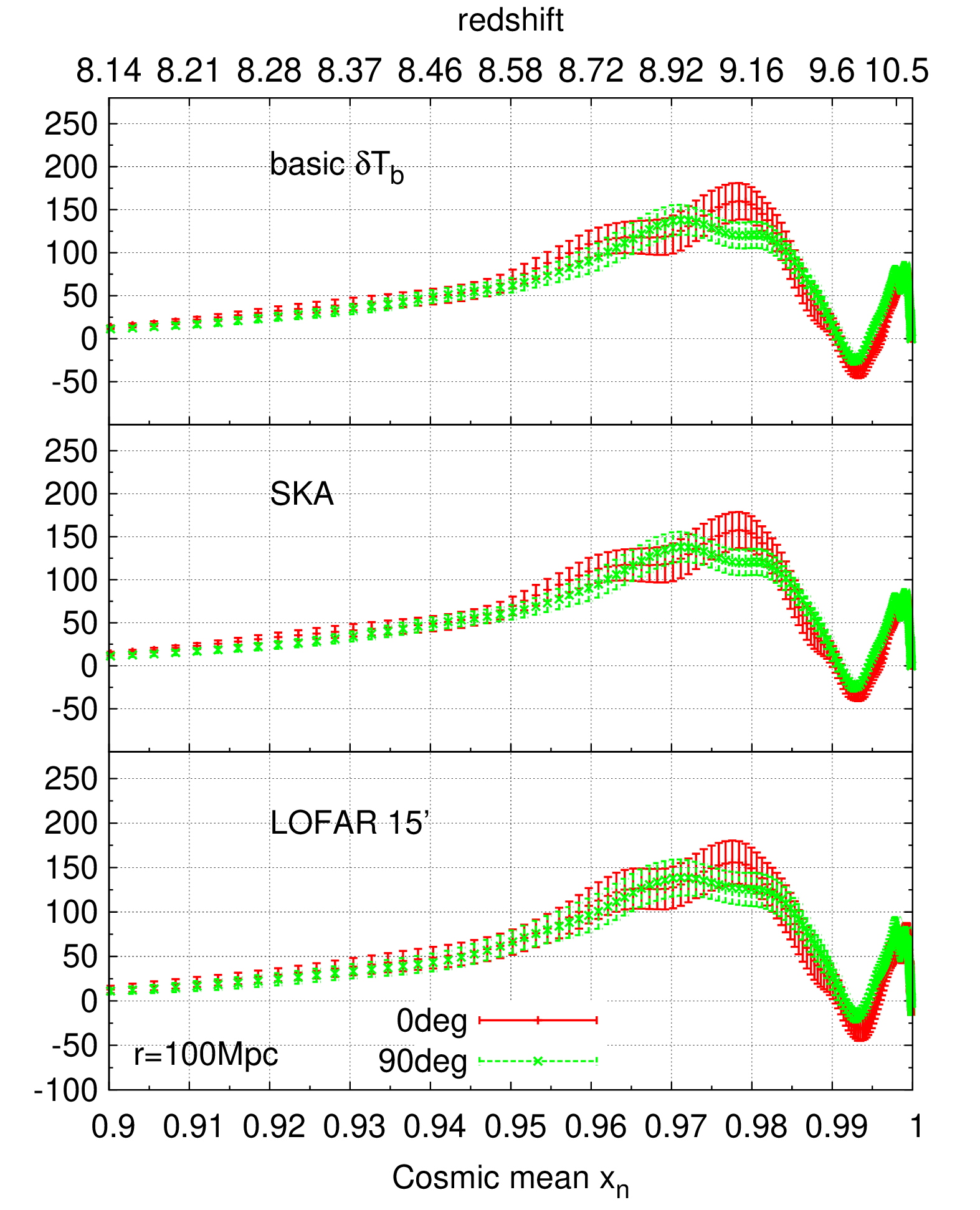}}
\caption{Two-point correlation functions calculated at a comoving separation of 100~cMpc along (red, solid) and perpendicular (green, dotted) to  the LOS for three different sets of data (from top to bottom): basic $\delta T_{\mathrm b}$, $\delta T_{\mathrm b}$+SKA and $\delta T_{\mathrm b}$+LOFAR 15'. Left panels show the range of the neutral fraction $0 \le {x_{\mathrm n}} \le 0.9$, and right panels $0.9 \le {x_{\mathrm n}} \le 1$. Basic $\delta T_{\mathrm b}$ and $\delta T_{\mathrm b}$+SKA are almost identical. Errors bars for the LOFAR case come from sample variance and instrumental noise. }
\label{corr_dTb_SKA_Lofar_0.9-1}
\end{figure*}

Figure~\ref{corr_dTb_SKA_Lofar_0.9-1}  shows how the estimates of the correlation along and perpendicular to the line of sight are affected. In this figure,
the error bars combine contributions from sample variance and instrumental noise. We could have estimated the latter by 
generating many realizations of the noise and computing the rms for each point. This would have been costly in CPU time. Instead we 
used an analytical estimate described in Appendix A. It is obvious that the instrumental noise from the SKA will not be an issue: it is small compared
to the level of sample variance that can be expected for a $20^\circ \times 20^\circ$ survey (36 times the size of our light cones). The size of the survey will be the critical 
parameter since the minimal value for the survey size considered in the current design is $5^\circ \times 5^\circ$. For the latter value, the anisotropic features of the correlation function may not be detected with statistical significance over the sample variance. In the case of LOFAR, the contribution of instrumental noise is similar to the sample variance for a $20^\circ \times 20^\circ $
 light cone. As a result, none of the anisotropy features are clearly detected. Reducing sample variance with a very large survey area might only result in a $2\sigma$ detection.
Moreover, limited resolution simulations tend to delay the beginning of 
reionization and the absorption regime anisotropies may well occur at higher redshifts than in our simulations. Consequently, the anisotropies in the absorption regime will probably not fall in the LOFAR band anyway.

\section{Conclusions}

To evaluate the amplitude of light cone anisotropies in the $21$~cm signal from the EoR, we run a suite of simulations in a 400~$h^{-1}$~cMpc box with $2 \times 512^3$ particles. Our simulations include dynamics (gravitation and hydrodynamics), source formation recipes, ionizing UV radiative transfer, X-ray radiative transfer and Lyman-$\alpha$
radiative transfer: this allows us to model the signal both in the absorption and in the emission regime. 

To quantify the anisotropy we examine the two-point correlation function of the differential brightness temperature of the $21$~cm signal perpendicular and parallel to the
line of sight, on scales of 20, 50 and 100~cMpc. Our first conclusion is that sample variance affects these correlation functions   
over much larger volumes than is usual for 
cosmological diagnostics. We find that a light cone $ > 10^\circ$ across  ($\sim 1.5$ comoving Gpc at $z \sim 10$) is necessary to reach statistically significant results. This is
due to the directional nature of our diagnostics. We are able to mitigate variance with our 400~$h^{-1}$~cMpc simulations by building several semi-independent light cones.

Using a sufficiently large sample size
we are able to find an anisotropy in the emission regime, but only on a 100~cMpc scale. This matches the theoretical prediction by \citet{Barkana06}.
While the amplitudes of our correlation functions are also in broad agreement with their predictions, we do not agree on the detailed behaviour. We find that the parallel correlation generally dominates over the perpendicular correlation at neutral fractions in the range $x_{\mathrm n}=0.2$ -- $0.5$. We interpret this as ionized bubbles connecting preferentially along the LOS during the
overlap phase.

Our analysis also includes the absorption regime. At the early stages of the EoR the correlation function parallel to the LOS rises above the perpendicular correlation at $x_{\mathrm n} \approx 0.98$ and the opposite is true in 
the $x_{\mathrm n} \approx 0.994$ -- $0.999$ range (the exact values depend on the nature of the sources). The feature at $x_{\mathrm n} \approx 0.98$ corresponds to the onset of heated regions. We interpret the anisotropic feature in the $x_{\mathrm n} \approx 0.994-0.999$ range as the onset of coupled regions. These regions are elongated along the light cone.

We checked that, on 100~cMpc scales, the inclusion of the velocity gradient contribution to $\delta T_{\mathrm b}$ does not change the shape of the correlation functions much. The velocity 
gradient contribution only moderately enhances the amplitude, well below the amplitude difference of the perpendicular and parallel correlation when the light cone anisotropy peaks. This is important since the velocity gradient contribution is also anisotropic and could confuse the light cone anisotropy signal.

We assess the observability of these anisotropic features by SKA and LOFAR using standard survey parameters. We assume that the residuals from foreground subtraction are much smaller than the signal. We find that SKA can detect all the anisotropy peaks we have identified, provided that the size of the survey is larger than $10^\circ \times 10^\circ$. In the case
of LOFAR, the emission regime anisotropy peak could only be marginally detected in a $15$~arcmin resolution survey with a very large size.

There are several reasons why it will be interesting to search for light cone anisotropy in the observations. First, the anisotropy peaks are connected to milestones in the
global reionization history: the onset of coupling and heating, and the overlap of ionized bubbles. The global history can only be indirectly retraced by interferometers that do not
give any information about the average signal. Second, the light cone anisotropy is characteristic of the signal and should not show the same features in the residual of
foreground subtraction. This could provide a test to check whether foregrounds have been correctly removed.

\section*{Acknowledgements}

The authors wish to thank Garrelt Mellema and Boudewijn Roukema for a thorough critical reading of the manuscript and the anonymous referee for a useful report. Part of this work consists of research conducted in the scope of the HECOLS International Associated Laboratory. Some of this work was carried out in the context of the LIDAU project. 
The LIDAU project is financed by the ANR (Agence Nationale de la Recherche, France) grant ANR-09-BLAN-0030.

\bibliographystyle{mn2e}
\bibliography{./myref}{}

\appendix

\section[]{Contribution of instrumental noise to errors bars of the 2-point correlation function}

Let us call $\delta T_{\mathrm b}({r_i},z)$ the differential brightness temperature of the $21$~cm signal of a pixel in the light cone at redshift $z$, and 
consider position ${r_i}$ in the plane perpendicular to the line of sight. The index $i$ fully covers the two-dimensional transverse maps. Then, let us
write $n({r_i},z)$ for the instrumental noise in the same pixel. In a given transverse map, $n$ is a realization of a Gaussian random field with
standard deviation $\sigma_n(z)$.

Let us define the correlation function perpendicular to the line of sight:
\begin{displaymath}
\xi(\Delta r,z) = {1 \over N} \sum\limits^N_{j=1}{ \left[ \delta T_{\mathrm b}({r_{1,j}},z) - \overline{ \delta T_{\mathrm b}}(z)\right]  \left[ \delta T_{\mathrm b}({r_{2,j}},z) - \overline{\delta T_{\mathrm b}}(z)\right]},
\end{displaymath}
where the summation extends over all pairs of pixels in the transverse map such that $|{r_{1,j}}-{r_{2,j}}|=\Delta r$. $\overline{ \delta T_{\mathrm b}}(z)$ is the average of the signal on the transverse map at redshift $z$.

The correlation function of the noised signal then reads:
 \begin{eqnarray*}
\xi_n(\Delta r,z)={1 \over N} \sum\limits_{j=1}^N{ \left[ \delta T_{\mathrm b}({r_{1,j}},z) + n({r_{1,j}},z) - \overline{\delta T_{\mathrm b}}(z) - \overline{ n}(z)\right] }\\
\qquad  \times \left[ \delta T_{\mathrm b}({r_{2,j}},z) + n({r_{2,j}},z) - \overline{ \delta T_{\mathrm b}}(z) - \overline{ n}(z) \right] 
\end{eqnarray*}

Due to the finite number of pixels, $ \overline{ n}(z)$ is non-zero. Its value is a realization of a Gaussian distribution function with standard deviation $\sigma_n(z) \over M$ where $M^2$ is the total number of independent pixels in the transverse map at redshift $z$ (Central limit theorem). If the transverse maps are several pixels thick, then $M$ increases. $M$ is set by the instrumental resolution.

Expanding the previous equation gives:
\footnotesize
\begin{eqnarray*}
\xi_n(\Delta r,z) \!& = &\!\!\xi(\Delta r,z) \\
&& \!\!+{1 \over N} \sum\limits_{j=1}^N \left[ \delta T_{\mathrm b}({r_{1,j}},z) - \overline{ \delta T_{\mathrm b}}(z)\right] \left[  n({r_{2,j}},z) - \overline{n}(z)\right]\\
&& \!\!+{1 \over N} \sum\limits_{j=1}^N \left[  n({r_{1,j}},z) - \overline{ n}(z)\right] \left[ \delta T_{\mathrm b}({r_{2,j}},z) - \overline{\delta T_{\mathrm b}}(z)\right] \\
&& \!\!+{1 \over N} \sum\limits_{j=1}^N  \left[  n({r_{1,j}},z) - \overline{ n}(z)\right]  \left[  n({r_{2,j}},z) - \overline{n}(z)\right]
\end{eqnarray*}
\normalsize

The second and third terms on the right-hand side are cross correlations of the signal and the noise, and the last term is the noise auto-correlation. Once again, even though the noise is uncorrelated and these terms would be zero in the limit of infinite sampling, they leave residuals  for the finite number $N$ of pixel pairs that contribute to the computation of the correlation function. Let us further expand and reorder this expression.

\begin{eqnarray*}
\xi_n(\Delta r)& = &\xi(\Delta r)\\
&& -{\overline{n}(z)\over N} \sum\limits_{j=1}^N \left[\delta T_{\mathrm b}({r_{1,j}},z) - \overline{\delta T_{\mathrm b}}(z) \right] \\
&& -{\overline{n}(z)\over N} \sum\limits_{j=1}^N \left[\delta T_{\mathrm b}({r_{2,j}},z) - \overline{\delta T_{\mathrm b}}(z) \right]\\
&& +{1 \over N} \sum\limits_{j=1}^N \left[ \delta T_{\mathrm b}({r_{1,j}},z) - \overline{\delta T_{\mathrm b}}(z)\right] n({r_{2,j}},z) \\
&& +{1 \over N} \sum\limits_{j=1}^N \left[ \delta T_{\mathrm b}({r_{2,j}},z) - \overline{\delta T_{\mathrm b}}(z)\right] n({r_{1,j}},z) \\
&& +{1 \over N} \sum\limits_{j=1}^N  n({r_{1,j}},z) n({r_{2,j}},z)  \\
&& + \overline{n}^2(z) -2\overline{n}(z)\langle  n\rangle_{\mathrm{pairs}}(z) 
\end{eqnarray*}
 
Here, $\langle\,\rangle_{\mathrm{pairs}}$ designates a weighted average over the transverse map. Each pixel contributes with a weight proportional to the number
of pairs it participates in: pixels in the centre, unaffected by the edges of the map (farther away than $\Delta r$) contribute more. On the other hand, the
weighting is uniform if we consider periodic boundary conditions; then $\overline{x} =\langle x \rangle_{\mathrm{pairs}}$. This question does not arise for the computation of the correlation function parallel to the line-of-sight. We will make the assumption $\overline{x} =\langle x \rangle_{\mathrm{pairs}}$. Thus,
the terms on the second and third line of the equation vanish.

If we write $\xi_n(\Delta r) = \xi(\Delta r) +\Delta \xi$ where $\Delta \xi$ is the deviation of the correlation function of the noised signal from the correlation function of the pre-noised signal at redshift $z$, then we can write:

\begin{eqnarray*}
\Delta \xi& = &{1 \over N} \sum\limits_{j=1}^N \left[ \delta T_{\mathrm b}({r_{1,j}},z) - \overline{\delta T_{\mathrm b}}(z)\right] n({r_{2,j}},z) \\
&& +{1 \over N} \sum\limits_{j=1}^N \left[ \delta T_{\mathrm b}({r_{2,j}},z) - \overline{ \delta T_{\mathrm b}}(z)\right] n({r_{1,j}},z) \\
&& +{1 \over N} \sum\limits_{j=1}^N  n({r_{1,j}},z) n({r_{2,j}},z)   - \overline{n}^2(z) \\ 
\end{eqnarray*}

The expectation value of the terms on the first two lines of the right-hand side (in the sense of averaging over many realizations of the noise) is zero. Since
$n$ and $\delta T_{\mathrm b}$ are uncorrelated:

$$
\mathrm{Dev}\left[{1 \over N} \sum\limits_{j=1}^N  \left[ \delta T_{\mathrm b}({r_{1,j}},z) - \overline{\delta T_{\mathrm b}}(z)\right] n({r_{2,j}},z)  \right] = { \sigma_{T_{\mathrm b}}(z) \sigma_n(z) \over M^2}
$$

where $\sigma_{T_{\mathrm b}}(z)$ is the standard deviation of $\delta T_{\mathrm b}$ in the 2D map at redshift $z$.

Since $n({r_j^1},z)$ and $n({r_j^2},z)$ are independent, we can write:
$$
\mathrm{Dev}\left[{1 \over N} \sum\limits_{j=1}^N  n({r_j^1}) n({r_j^2}) - \langle  n\rangle^2\right] = 2{ \sigma^2_n(z) \over M^2} . 
$$

\begin{figure}
\includegraphics[width=0.7\linewidth, angle=-90]{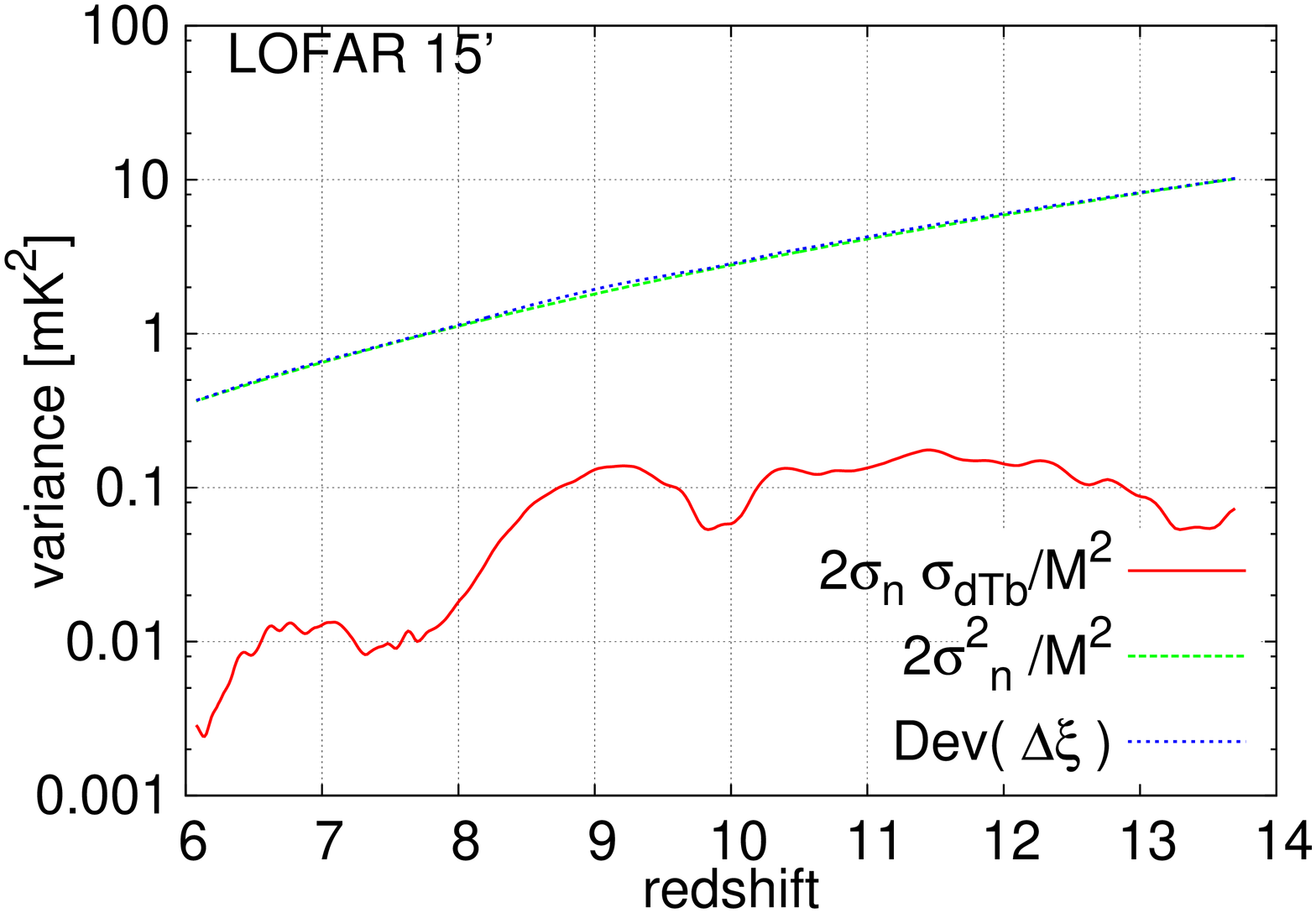}
\caption{ Contribution of instrumental noise to errors bars of the 2-point correlation function. }
\label{instrumental_noise}

\end{figure}

Finally, we get the following expression for the standard variation of the correlation function (error bar due to noise):
$$
\mathrm{Dev}( \Delta \xi)=  2  { \sigma_n(z) \sigma_{T_{\mathrm b}}(z)\over M^2} + 2  { \sigma^2_n(z) \over M^2}
$$
In observations it would not be possible to evaluate $\sigma_{T_{\mathrm b}}(z)$. But in our case we can compute it from the simulated signal.
The relative contribution of each term is plotted in Fig.~\ref{instrumental_noise}.

\newpage

\bsp

\label{lastpage}

\end{document}